\documentclass[sigconf]{acmart}
\AtBeginDocument{%
  \providecommand\BibTeX{{%
    \normalfont B\kern-0.5em{\scshape i\kern-0.25em b}\kern-0.8em\TeX}}}

\copyrightyear{2021}
\acmYear{2021}
\setcopyright{rightsretained}
\acmConference[RAID '21]{24th International Symposium on Research in Attacks, Intrusions and Defenses}{October 6--8, 2021}{San Sebastian, Spain}
\acmBooktitle{24th International Symposium on Research in Attacks, Intrusions and Defenses (RAID '21), October 6--8, 2021, San Sebastian, Spain}
\acmDOI{10.1145/3471621.3471849}
\acmISBN{978-1-4503-9058-3/21/10}

\usepackage{amsfonts}
\usepackage{algorithmic}
\usepackage{textcomp}
\usepackage{xcolor}
\usepackage{lipsum}
\usepackage{breakurl}
\usepackage{subcaption}
\usepackage{multirow}
\usepackage{float}
\usepackage{color, colortbl}
\usepackage{listings}
\usepackage{enumitem}

\newcounter{rownumber}
\definecolor{codegreen}{rgb}{0,0.6,0}
\definecolor{codegray}{rgb}{0.5,0.5,0.5}

\def\System{IskiOS}
\def\Comment#1{}
\def\BibTeX{{\rm B\kern-.05em{\sc i\kern-.025em b}\kern-.08em
    T\kern-.1667em\lower.7ex\hbox{E}\kern-.125emX}}

\begin{document}

\title{Fast Intra-kernel Isolation and Security with {\System}}

\author{Spyridoula Gravani}
\orcid{0000-0002-9221-5089}
\author{Mohammad Hedayati}
\orcid{0000-0002-4036-3354}
\email{{sgravani,hedayati}@cs.rochester.edu}
\affiliation{%
  \institution{University of Rochester\raisebox{-1ex}{\strut}}
  \country{}}

\author{John Criswell}
\author{Michael L. Scott}
\orcid{0000-0001-8652-7644}
\email{{criswell,scott}@cs.rochester.edu}
\affiliation{%
  \institution{University of Rochester}
  \country{}}

\begin{abstract}
The kernels of operating systems such as
Windows, Linux, and \mbox{MacOS} are vulnerable to control-flow hijacking.
Defenses exist, but many require efficient intra-address-space
isolation.  Execute-only memory, for example, requires read protection
on code segments, and shadow stacks require protection from
buffer overwrites.
Intel's Protection Keys for Userspace (PKU) could, in principle,
provide the intra-kernel isolation needed by such defenses,
but, when used as designed, it applies only to user-mode application code.

This paper presents an unconventional approach to memory protection,
allowing PKU to be used within the operating system kernel on existing Intel hardware,
replacing the traditional user/supervisor isolation mechanism and,
simultaneously, enabling efficient intra-kernel isolation.  We call the
resulting mechanism \emph{Protection Keys for Kernelspace} (PKK).
To demonstrate its utility and efficiency, we present a system we call
{\System}: a Linux variant featuring execute-only memory (XOM)
and the \emph{first-ever race-free shadow stacks for x86-64}.

Experiments with the LMBench kernel microbenchmarks display a geometric mean
overhead of about 11\% for PKK and no additional overhead for XOM\@.
\System's shadow stacks bring the total to 22\%.
For full applications, experiments with the system benchmarks of
the Phoronix test suite display negligible overhead for PKK and XOM,
and less than 5\% geometric mean overhead for shadow stacks.
\end{abstract}



\maketitle

\section{Introduction}
\label{section:intro}
As the operating system (OS) kernel forms the foundation of the software stack,
control-flow hijacking attacks~\cite{ROP:TOISS12} on the OS kernel
jeopardize the security of the entire system.
To safeguard system integrity,
\emph{control-flow integrity (CFI) enforcement}~\cite{KCoFI:Oakland14,
  KCFI:EuroOakland16, DROP-ROP:Blackhat17, Fine-CFI:TIFS18}
ensures that only paths in an authorized control-flow graph
are followed during execution~\cite{CFI:TISSEC09}.
At the same time,
\emph{leakage-resilient diversification}~\cite{KernelDiversification:CNS16,
  kRX:TOPS19, Readactor:SP15, LR2:NDSS16, HideM:CODASPY15, XnR:CCS14}
undermines the attacker's knowledge of the code layout in memory,
making successful exploitation more difficult.
Both techniques must efficiently isolate memory within the OS kernel.
CFI must protect the integrity of return addresses on the
stack~\cite{ControlFlowBending:UsenixSec15, OutOfControl:Oakland14,
  ControlFlowBending:UsenixSec15, RTC-CFI:ACSAC2018,
  CFIEvaluation:CSUR17, LosingControl:CCS15};
diversification must prevent attackers from learning the layout of
kernel code---e.g., by leveraging buffer
overreads~\cite{Overread:EUROSEC09,JITROP:SP13}.

Existing intra-kernel isolation mechanisms generally rely on
\emph{software fault isolation} (SFI)~\cite{SFI:SOSP93,
  SFI2:UsenixSec10}, adding bounds checks to memory
writes~\cite{KCoFI:Oakland14} and reads~\cite{kRX:TOPS19} to prevent
unauthorized access to sensitive data.  As the overhead of such checks is
proportional to the number of protected regions,
SFI-based approaches often group all sensitive information
together---typically
at the upper end of
the virtual address space---so only one bound must be
checked~\cite{kRX:TOPS19, KCoFI:Oakland14}. Unfortunately, this
grouping reduces design flexibility and
diversification entropy.
The most efficient SFI implementation on \mbox{x86-64}~\cite{kRX:TOPS19} relies
on Intel's Memory Protection Extensions (MPX), a hardware feature that
has recently been deprecated~\cite{MPXDeprecated}.

In this work, we introduce an alternative approach to intra-kernel
memory isolation based on novel use of Intel's Protection Keys for
Userspace (PKU)~\cite{IntelArchManual21}.
PKU allows each page table entry (PTE) to specify one of
16 protection keys (PKEYs); at run time, unprivileged software can
dynamically disable read access or both read \& write access to pages on
a PKEY-by-PKEY basis.
However, as the name implies, PKU applies only to pages whose
PTEs are marked as user space (i.e., accessible to application code).

We have developed a new user/kernel isolation
mechanism, dubbed \emph{PKK (Protection Keys for Kernelspace)},
which uses PKU hardware to dynamically disable read and/or
write access to pages while running in the OS kernel.
Our PKK mechanism is both inexpensive and flexible, and supports
both SMEP and SMAP (Supervisor-Mode Execution/Access
Prevention~\cite{IntelArchManual21}).  Additionally, while it is
fully compatible with Kernel Page Table Isolation (KPTI)~\cite{KPTI:LKML},
PKK \emph{does not require} KPTI in order to function, allowing systems that
do not need or do not want KPTI to avoid its overhead.\footnote{
    As part of its Trust Domain Extensions (TDX), Intel has recently
    announced that certain future processors will provide hardware
    support for Protection Keys in Supervisor mode (PKS)\@.  PKS
    machines will allow 16 protection keys to be used in supervisor mode
    in addition to the 16 in user mode.  Our PKK mechanism provides
    PKS-like functionality on existing machines.
}

To demonstrate the effectiveness of PKK, we developed
\emph{{\System}},\footnote{``{\System}'' means ``shadow'' in Greek.}
a system that provides execute-only memory (XOM) for storing code and
write-protected shadow
stacks for the Linux kernel.  Unlike previous
approaches~\cite{kRX:TOPS19}, {\System} protects both the code and
the shadow stacks efficiently, with no changes to virtual address space
layout; \System\ is thus compatible with existing kernel memory
allocators and imposes no limits on the entropy of diversification strategies.

To protect the shadow stack, \System\ incorporates a novel calling
convention
that is immune not only to control-flow hijacking attacks in the calling
thread but also to attacks that leverage cross-thread
races~\cite{RFG:MSRC, Clang10ShadowCallStack:Web}.
Along with this calling convention, we
introduce a pair of techniques
to reduce the frequency of PKU domain switches.
The first technique, dubbed \emph{shadow write optimization}, avoids
redundant writes of return addresses to the shadow stack.
The second performs aggressive function inlining, avoiding calls to many
smaller functions altogether.

Our contributions can be summarized as follows:

\begin{itemize}
  \item
    We demonstrate that Intel PKU can be used to enable efficient user/kernel
    isolation, to widen the set of protection modes for kernel memory, and to
    change protections cheaply at fine temporal granularity.

  \item
    We describe a system, {\System}, that leverages the resulting PKK
    mechanism to provide execute-only memory (XOM) and protected shadow stacks
    within the Linux kernel, with arbitrary
    address space layout.  To the best of our knowledge, ours is
    the \emph{first write-protected, race-free shadow stack for the x86-64}.

  \item
    We describe a \emph{shadow write optimization} which, together with
    aggressive inlining, serves to significantly reduce the
    remaining overheads of PKK-based shadow stacks.

  \item
    Using the LMBench microbenchmark suite~\cite{lmbench}, we
    demonstrate a geometric mean overhead, relative to standard Linux,
    of approximately 11\% for PKK and XOM\@.  Adding our protected
    shadow stacks brings the total to roughly 22\%.

  \item
    Using workloads drawn from performance regression tracking of the
    mainline Linux kernel~\cite{benchmark:PTS:LinuxRegression}), we
    confirm that overheads are substantially lower for full-size programs:
    the average overhead of PKK and XOM becomes negligible, and
    protected shadow stacks add less than 5\% geomean overhead.
\end{itemize}
Source code for {\System} is available at
\url{https://github.com/URSec/iskios}.

\section{Background}
\label{section:back}
\subsection{User/Kernel Isolation on x86}
\label{section:back:isolation}

For efficiency, Linux has traditionally mapped kernel code, data, and
loadable modules into the address space of every running process.  This
convention allows system calls, traps, and interrupts to vector into the
kernel without flushing entries
from the TLB\@.  It also allows the kernel to easily access
the active process's user-level memory.

To support this address space sharing, the x86 includes a
User/\linebreak[1]Supervisor (U/S) flag in every page table entry
(PTE)~\cite{IntelArchManual21}.  Pages in the
upper region of the address space, which map kernel code and data, are marked as
supervisor-only: accessing such a page while running in user mode
(Ring 3) results in a trap~\cite{IntelArchManual21}.\footnote{%
    To mitigate
    Meltdown~\cite{Meltdown:UsenixSec18} attacks, some recent versions
    of Linux have employed separate page tables for user and supervisor
    code (i.e., KPTI~\cite{KPTI:LKML}).  The resulting security comes at
    a significant performance cost~\cite{KPTI-cost}; long-term solutions
    depend on hardware updates that restore the viability of user/kernel
    address space sharing.}

\subsection{Protection Keys for Userspace}
\label{section:back:pku}

In its Skylake generation of processors, Intel introduced
\emph{Protection Keys for Userspace} (PKU)~\cite{IntelArchManual21}.
A descendant of mechanisms dating back to the IBM 360~\cite{360-PoO},
PKU is unusual in that it applies only to user-mode pages and allows
permissions to be changed by an unprivileged instruction.
While PKU is intended mainly as a memory safety enhancement,
researchers have used
it~\cite{Hodor:ATC19,ERIM:UsenixSec19,SoKShiningLight:SP19,Memcached:ICPP20} to provide
intra-process isolation for user-space applications.

PKU introduces a 32-bit register called {\tt pkru} and instructions
({\tt rdpkru} and {\tt wrpkru}) to read and write
it~\cite{IntelArchManual21}.
Four previously unused bits (bits 62:59) in each PTE
are then used to associate the page with one of 16 possible protection
domains.  The {\tt pkru} uses two bits per key---dubbed access disable
(AD) and write disable (WD)---to encode
access rights that are restricted in each domain.
On any load or store to a user-space page, the processor checks
permissions as usual and then drops all rights (if any) associated
with the page's protection key in the {\tt pkru} of
the currently executing hyperthread.
The processor ignores the protection key for instruction fetches and for
loads and stores of addresses whose PTE indicates kernel space.
Thus, in the expected use case, PKU does not affect accesses to kernel
code or data.

To date, intra-address-space isolation in the OS kernel has relied on
SFI~\cite{SFI:SOSP93,SFI2:UsenixSec10},
which adds instrumentation (e.g., bit-masking
operations~\cite{KCoFI:Oakland14} or MPX
bounds checks~\cite{Apparition:UsenixSec18,kRX:TOPS19}) before stores and/or
loads.
Enabling protection keys in the kernel would bring
two key benefits that are difficult or impossible to attain with
existing SFI-based approaches.

First, PKU provides more flexibility in address space layout.
To improve performance, solutions using SFI~\cite{SFI:SOSP93} often place
pages of the same protection domain in contiguous virtual
addresses~\cite{KCoFI:Oakland14, kRX:TOPS19, Datashield:ASIACCS17,
  Apparition:UsenixSec18}.
This approach requires significant engineering effort
(for memory allocators in particular~\cite{kRX:TOPS19})
and reduces the entropy of code-layout randomization schemes.
By contrast, protection keys permit pages in different domains to be
located anywhere within the virtual address space with no
performance or entropy loss.

Second, PKU easily supports up to 16 protection
domains~\cite{IntelArchManual21}.
By contrast, the complexity of bit-masking bounds checks
increases with the number of protection domains~\cite{KCoFI:Oakland14},
and Intel~MPX~\cite{IntelArchManual21}
(which is now deprecated~\cite{MPXDeprecated})
provides only four bounds registers.
With PKU, a single \texttt{wrpkru} instruction can change the access
rights of an arbitrary subset of the 16 protection domains.

\section{Protection Keys for Kernelspace}
\label{section:pkk}
Our goal is to enable what one might call \emph{Protection Keys for
Kernel\-space (PKK)} while maintaining the traditional guarantees of
user/kernel isolation.
In a conventional system, however, kernel code and data are placed in
pages for which the U/S flag is clear (indicating they can be accessed
only in supervisor mode), and
the hardware ignores the PKU bits when accessing such pages.
Fortunately, we observe that we can use memory protection keys
to subsume the functionality of the U/S bit, and then leave that bit
set in the PTEs of both user and kernel pages.

There are only two things that the OS kernel can
do that user code cannot: execute privileged instructions and access
protected OS kernel memory.  The execution of privileged instructions is
controlled not by the U/S bit in PTEs but by the processor's
\emph{execution mode} (Ring 3 v.\ Ring 0)~\cite{IntelArchManual21},
which we do not change.
The effect of the U/S bit on loads and stores can then be emulated by
assigning user and supervisor pages to different protection domains,
enabling and disabling access to these domains when crossing from
user to kernel space or vice versa, and preventing changes to {\tt pkru}
from occurring at other times (see Section~\ref{section:pkk:safety} below).
As \texttt{pkru} permissions have no impact on instruction fetches, code
running in user-mode can jump directly to code within the kernel's
code segment.  However, such a jump into kernel code is harmless
\emph{because the processor is still in user-mode (Ring~3)}:
executing a privileged instruction causes a trap (because the execution mode
is still unprivileged), and reading or writing kernel
memory causes a trap (because access is locked out by the {\tt pkru}).
As now, the only way to enter the kernel is to use a system call, trap, or
interrupt which will change the processor's mode from user-mode to
kernel-mode.

To implement PKK,
{\System} sets the U/S bit in every entry of every page table, with the
exception of a few entries used to map trampoline pages that handle system
calls and interrupts.  This effectively marks all memory as user mode.
{\System} then reserves keys \mbox{0--7} for use as \emph{kernel protection keys}.
Every page of kernel memory will be assigned one of these eight keys;
{\System} disables both read and write access to pages with these keys
when the system is operating in user mode and enables them (selectively,
as dictated by the security policy) when executing kernel code.
Existing applications that make legitimate use of PKU can continue
to obtain keys from the kernel through the
\texttt{pkey\_alloc()}~\cite{Linux:pkeysyscalls} system call,
which is modified to return an available key between 8 and
15. As applications may use PKEYs 8--15 to enforce their
own security policies, {\System} saves the value of the \texttt{pkru}
on kernel entry and restores it on return to user space.

\subsection{Controlling Use of \texttt{wrpkru}}
\label{section:pkk:safety}

\begin{center}
\begin{minipage}[c]{0.45\textwidth}
\centering
\lstset{
  belowcaptionskip=0.8\baselineskip,
  breaklines=true,
  xleftmargin=2em,,
  language=C,
  showstringspaces=false,
  numberstyle=\color{codegray},
  numbers=left,
  numbersep=7pt,
  basicstyle=\footnotesize\ttfamily,
  keywordstyle=\bfseries\color{green!40!black},
  commentstyle=\itshape\color{purple!40!black},
  identifierstyle=\color{blue},
  stringstyle=\color{orange},
  escapeinside={{*@}{@*}},
  morestring=[b]",
}
\begin{lstlisting}[caption={glibc Modification.},
label={lst:glibc-wrpkru}]
/* Overwrite the PKRU register with VALUE.  */
static inline void pkey_write(unsigned int value) {
  __asm__ volatile(
      "Lmask: \n\t" *@\label{glibc-Lmask}@*
      "orl  $0x0000ffff, *@\color{orange}\%@**@\color{orange}\%@*eax \n\t" *@\label{glibc-mask16}@*
      "wrpkru \n\t" *@\label{glibc-wrpkru-inst}@*
      "cmpw  $0xffff, *@\color{orange}\%@**@\color{orange}\%@*ax \n\t" *@\label{glibc-hodor-test}@*
      "jne Lmask \n\t" *@\label{glibc-jump}@*
      :: "a"(value), "c"(0), "d"(0));
}
\end{lstlisting}
\end{minipage}
\end{center}

Since \texttt{wrpkru} is not a
privileged instruction ~\cite{IntelArchManual21}, application code can modify
the \texttt{pkru} at will.  {\System}
must therefore prevent application code from enabling read or write
access to kernel pages. To do so, {\System} must
address two challenges.  First, it must allow applications to
use the \texttt{wrpkru} instruction to enforce their own security policies
while preventing them from clearing the WD or AD bits for PKEYs 0--7.
Second, given that (unprivileged) kernel instructions are now executable
in user mode, \System\ must prevent applications from compromising security
by jumping to \texttt{wrpkru} instructions in the kernel.

\textbf{Restricting Application wrpkrus}
To address the first challenge, {\System} borrows techniques from
our earlier Hodor project~\cite{Hodor:ATC19}: {\System} ensures that all
application pages with execute permission have write permission disabled;
when the kernel maps a page into the virtual
address space of an application with execute permissions, it scans the page
(including boundaries with the pages before and after it) for any byte
sequence that might comprise a \texttt{wrpkru} instruction.  If it finds any,
it places a debug watchpoint~\cite{IntelArchManual21} on the address. If the
address is ever used for an
instruction fetch, the watchpoint generates a trap. {\System} then inspects
the value of the intended write to the \texttt{pkru} to ensure that the
process is not modifying PKEYs \mbox{0--7} and, if it is,
terminates the process. This convention allows applications to continue
using PKEYs \mbox{8--15} without restriction.
Programs that generate code dynamically can
write code to a page and then use {\tt mprotect()} to change the
permissions from writable and non-executable to non-writable and executable;
{\System} will scan the page when changing the permissions.

\begin{center}
\begin{minipage}[c]{0.45\textwidth}
\lstset{
 language=[x86masm]Assembler,  
 morekeywords={CDQE,CQO,CMPL,CMPB,CMPW,CMPSQ,CMPXCHG16B,JRCXZ,JMPQ,ADDQ,LODSQ,MOVSXD,%
                  POPFQ,PUSHFQ,SCASQ,STOSQ,IRETQ,RDTSCP,SWAPGS,TESTB,UD2,LEAQ,
                  CALLQ,BT,JB, %
                  VMFUNC,WRPKRU,XORL,RDPKRU,MOVL,MOVQ,LFENCE,RETQ, %
                  rax,rdx,rcx,rbx,rsi,rdi,rsp,rbp,rip, %
                  r8,r8d,r8w,r8b,r9,r9d,r9w,r9b, %
                  r10,r10d,r10w,r10b,r11,r11d,r11w,r11b, %
                  r12,r12d,r12w,r12b,r13,r13d,r13w,r13b, %
                  r14,r14d,r14w,r14b,r15,r15d,r15w,r15b},
  belowcaptionskip=0.8\baselineskip,
  xleftmargin=2em,
  basicstyle=\footnotesize\ttfamily,
  keywordstyle=\color{black}\bfseries,
  commentstyle=\itshape\color{purple!40!black},
  numberstyle=\color{codegray},
  numbers=left,
  numbersep=7pt,
  breaklines=true,
  commentstyle=\color{codegreen},
  escapeinside={{*@}{@*}},
}
\begin{lstlisting}[caption={Safe \texttt{wrpkru} in Kernel.},
label={lst:kernel-wrpkru}]
; wrpkru instance in kernel code
wrpkru *@\label{safe-wrpkru}@*

; ensure execution is in kernel mode
movq 	*@\%@*cs, *@\%@*rcx *@\label{pkk-safety-load-cs}@*
testb	$3, *@\%@*cl   *@\label{pkk-safety-byte-test}@*
je   	Lskip 	  *@\label{pkk-safety-skip}@*
Ltrap:	*@\label{pkk-safety-Ltrap}@*
ud2  *@\label{pkk-safety-ud2}@*

Lskip:	*@\label{pkk-safety-Lskip}@*
; ensure expected permissions are set in pkru
cmpw $EXP_PERM, *@\%@*ax *@\label{pkk-safety-hodor-test}@*
jne Ltrap *@\label{pkk-safety-trap}@*
\end{lstlisting}
\end{minipage}
\end{center}

Our work on Hodor showed~\cite{Hodor:ATC19} that
existing Linux applications almost never contain more than one or two
\texttt{wrpkru} instructions, confirming that the debug watchpoint
approach is practical for existing applications.  One might worry that as
more applications come to use PKU, the four available watchpoint
registers~\cite{IntelArchManual21} might become a scarce resource.  We
observe, however, that applications making legitimate use of protection
keys can be expected to rely on the routines provided in Linux's C
Standard Library (\texttt{libc}) to change the \texttt{pkru}.
To avoid using debug watchpoints in the common case,
{\System} adds lines~\ref{glibc-Lmask}--\ref{glibc-mask16}
and~\ref{glibc-hodor-test}--\ref{glibc-jump} in Listing~\ref{lst:glibc-wrpkru}
to the \texttt{pkey\_write} library
function in GNU's \texttt{glibc}.  These mask the lower
16~bits of the value written to the \texttt{pkru} register.  With this change,
\texttt{glibc} requires no watchpoint because it will always
ensure that access to kernel pages is
disabled.\footnote{A signal could occur during \texttt{pkey\_write}, but
  {\System}
disables access to PKEYs \mbox{0--7} before dispatching signal
handlers and when resuming application execution after a signal
handler returns.}

\textbf{Protecting Kernel wrpkrus}
The second challenge reflects the fact that \texttt{wrprku}
instructions in the kernel typically \emph{will} clear the AD and WD
bits of kernel PKEYs.
To prevent application code from using these instructions,
{\System} inserts a check after each \texttt{wrpkru} in kernel code
to ensure that the processor is currently running in Ring~0, as
specified by the code segment register (\texttt{cs})---see
lines \mbox{\ref{pkk-safety-load-cs}--\ref{pkk-safety-skip}} in
Listing~\ref{lst:kernel-wrpkru}.
In addition, {\System} ensures that the expected permissions have been set,
thereby avoiding an attack in which
overly generous permissions are placed in
\texttt{eax} before jumping to a \texttt{wrpkru} instruction\footnote{%
    In a future system using PKS hardware, we would need similar checks
    to prevent the gadget-ization of \texttt{wrmsr} instructions, which
    can be used to disable PKS altogether.  Unlike \texttt{wrpkru},
    which impacts only the protection keys and is therefore relatively
    rare, \texttt{wrmsr} serves many purposes and appears in many parts
    of the OS kernel.}
(lines~\ref{pkk-safety-hodor-test}--\ref{pkk-safety-trap} in
Listing~\ref{lst:kernel-wrpkru}).
If any check fails, a production system would terminate the current process;
for simplicity, our prototype simply crashes
(line~\ref{pkk-safety-ud2} in Listing~\ref{lst:kernel-wrpkru}).

\subsection{SMEP Compatibility}
\label{section:pkk:smep}

Recent Intel processors provide Supervisor-Mode Execution
Prevention (SMEP)~\cite{IntelArchManual21} to harden
the kernel against \emph{ret2usr}~\cite{kGuard:UsenixSec12} attacks.
When SMEP is enabled, kernel-mode code cannot fetch instructions from pages
whose PTEs are marked as user space; any such access will cause a
page fault, allowing the OS to intervene.
Since {\System} configures kernel memory as user memory, we
must disable SMEP for kernel code to execute.
To prevent the kernel from executing arbitrary user code without SMEP support,
{\System} adopts kGuard's~\cite{kGuard:UsenixSec12} approach and adds
control-flow checks on kernel code. These checks, described in more
detail in Section~\ref{section:impl:llvm:smep}, apply to every indirect
control transfer and ensure that privileged execution remains within kernel
space. If execution attempts to cross to an address in the user portion
of the virtual address space, {\System} can intervene (our prototype
simply halts).

\subsection{SMAP Compatibility}
\label{section:pkk:smap}

Supervisor-Mode Access Prevention (SMAP)~\cite{IntelArchManual21}
is a CPU extension that disables supervisor-mode accesses to user
pages in an attempt to prevent attacker-controlled pointers from accessing
user memory directly, possibly subverting the kernel's control
flow~\cite{FunwithNULLptrs:LWN}. When the operating system needs to access
user memory for legitimate purposes
(e.g., \texttt{copy\_to/from\_user()}~\cite{UnderstandingLinux2:BovetCesati}),
it can temporarily
disable SMAP protection by setting the alignment checking flag bit in the
EFLAGS status register~\cite{IntelArchManual21}.
{\System} configures all linear addresses to be
user mode. Consequently, SMAP must be disabled for kernel code to be able to
access its own data. As with SMEP, {\System} clears the SMAP bit at boot
time.  To replicate SMAP's protections, {\System}
disallows kernel access to pages tagged with keys \mbox{8--15} (i.e.,
user pages) by default.  The
\texttt{copy\_to/from\_user()}~\cite{UnderstandingLinux2:BovetCesati}
functions temporarily enable access to pages tagged with keys
\mbox{8--15} when they need to read or write application memory.

\subsection{Side Channels}
\label{section:pkk:side}

PKK does not introduce any speculative execution side channels on processors
following the Skylake generation.\footnote{Skylake processors are susceptible
to both Meltdown and Meltdown-PK attacks, with or without PKK.}
Specifically, hardware mitigations against
Meltdown~\cite{Meltdown:UsenixSec18} guarantee that data will never be loaded,
even speculatively, if protection keys do not allow it~\cite{Intel-Security-Features}.
However, since PKK makes kernel code executable from user space, user programs
may hypothetically be able to infer the location of kernel code fragments by jumping
to random locations and observing the side effects on processor and memory
state.  To minimize the window of vulnerability, {\System} will terminate and
blacklist an application, regardless of installed signal handlers,
the first time it traps when executing instructions within the
kernel code segment in user-space.

\section{Enhancing Security with PKK}
\label{section:usecases}

We now describe our threat model, and
how we use Intel PKU to (1) make kernel
code
unreadable (execute-only), thereby preventing buffer
overreads~\cite{Overread:EUROSEC09} from revealing
potential gadgets, and (2) provide the kernel with
efficient, race-free shadow stacks.

\subsection{Threat Model}
\label{section:pkk:threat}

We assume the attacker aims to execute a
computation within the OS kernel with supervisor privileges.
The kernel itself is non-malicious but may
have exploitable memory safety errors such as buffer
overflows~\cite{AlephOne:StackSmash} and dangling
pointers~\cite{AfekSharabani:BlackHat07}.
Our attacker is an
unprivileged user: they can execute arbitrary code in user space but
cannot direct the OS kernel to load new kernel
modules implementing malicious code.\footnote{Systems such as
SecVisor~\cite{SecVisor:SOSP07} can prevent the loading of such
malicious code if privileged user-space tools cannot be trusted.}
We assume the enforcement of the W\^{}X~\cite{PAX:NOEXEC, IntelArchManual21,
AMDArchManual17} policy that prevents the attacker from injecting code directly
into kernel memory. Finally, we assume that the hardware is correctly
implemented and that side channels are out of scope.

\subsection{Kernel XOM}
\label{section:usecases:xom}

Code reuse attacks (e.g., ROP~\cite{ROP:TOISS12}) utilize information
on where and how code has been placed in memory.  To thwart such attacks,
code diversification schemes randomize the layout and location of code
so that attackers cannot locate useful gadgets and
functions using a priori knowledge~\cite{ASLP:ACSAC06, SmashingGadgets:SP12, Readactor:SP15}.
Advanced code reuse attacks (e.g., JIT-ROP~\cite{JITROP:SP13}) exploit
buffer overread bugs~\cite{Overread:EUROSEC09} to inspect the code
segment to locate reusable code.  To prevent such attacks,
{\System} places all kernel and kernel module code pages in
\emph{eXecute Only Memory~(XOM)}---memory that can be executed but not
read or written.
The x86 architecture does not support XOM directly, but {\System} implements
it using PKK.

Specifically, {\System} reserves one of the 8 OS kernel protection keys for
the OS kernel code segment.  It configures all page table entries for pages
containing kernel code to use this key.  It then sets the access
disable (\texttt{AD}) bit in the \texttt{pkru} for this key,
disabling read access to pages containing kernel code.
Since protection keys affect only data accesses, instruction fetch and
execution are permitted~\cite{IntelArchManual21}.
Unlike SFI-based implementations of XOM~\cite{kRX:TOPS19, LR2:NDSS16},
{\System} can protect code pages at any virtual address, placing no
restrictions on diversification: code pages and data pages can be
interspersed and placed wherever desired in the virtual address space.

{\System} also needs to protect \emph{physmap}, the direct map of
physical memory used in kernel space~\cite{ret2dir:UsenixSec14}.
Physmap facilitates dynamic kernel memory allocation, but it also
causes address aliasing---more than one virtual address mapping to the
same physical address. To prevent buffer overreads from reading the kernel code
segment through its mapping in physmap, {\System} ensures that both
mappings of the kernel code segment (including the one in physmap)
use the same reserved kernel protection key.

\subsection{Kernel Shadow Stack}
\label{section:usecases:shadow}

Advanced code-reuse attacks commonly modify return addresses on the
stack~\cite{AOCR:NDSS17, PIROP:EuroSP18}. A \emph{shadow stack} protects
against such attacks by keeping return addresses where they cannot
easily be modified.
While the location of the shadow stack could simply be obscured via
randomization~\cite{CPI:OSDI14}, such schemes have proven
insufficient~\cite{UnderminingIH:UsenixSec16, LosingControl:CCS15,
DualStack:ASIACCS18}.  For real protection, we need the shadow stack to
be \emph{inaccessible}.

We also need it to be \emph{race free}, with no timing windows during
which a return address can be modified by an attacker on another core.
Previous shadow stack implementations~\cite{RAD:ICDCS01,
  ROPdefender:ACIACCS11, ShadowStack:ASIACCS15, SoKShiningLight:SP19,
  RFG:MSRC} have instrumented function prologues
to copy the return address from the original stack to the shadow stack.
Such designs leave the return address vulnerable to modification by
another thread after the call instruction saves it
but before the function prologue copies it.  Several previous
schemes~\cite{RAD:ICDCS01, ROPdefender:ACIACCS11, RFG:MSRC} also exhibit
races in
function epilogues when they verify the validity of the
return address (or copy it from the shadow stack back to the original
stack) prior to executing a return instruction:
the return address can be corrupted after it is validated
(or copied from the shadow stack) but before the called function returns.
While the windows of vulnerability in these schemes are small,
real-world implementations
such as Microsoft's Return Flow Guard (RFG)~\cite{RFG:MSRC} and Clang's shadow
stack for x86-64~\cite{Clang10ShadowCallStack:Web} were removed quickly after
their public release as the inherent races were shown to be exploitable.

{\System} utilizes PKK, together with call-site
modification, to provide what we believe to be \emph{the first
  write-protected, race-free shadow stack for the x86}.
To avoid the need for a separate stack pointer,
we utilize a
\emph{parallel shadow stack}~\cite{ShadowStack:ASIACCS15} design in
which all shadow stack entries are located at a constant offset from their
locations on the original stack.
To protect the shadow stack from tampering, we dedicate one of the
kernel protection keys to pages used for shadow stacks
(and the pages that map to the same frames in physmap).
During normal execution, write
access to the shadow stack is disabled while read access remains
enabled.  When {\System} needs to write a copy of the return address
to the shadow stack, it
temporarily enables write access to the shadow stack's protection key
(PKEY~2) in the \texttt{pkru} (this impacts only the current hardware
thread), writes the return address to the
shadow stack, and then revokes write access.  Function return requires
no changes to the \texttt{pkru} as the shadow stack is always
readable.

To provide race-freedom, {\System} modifies the default calling convention
to pass the return address to each function in a register, rather than
pushing it onto the stack.  We modify the code generator to reserve
\texttt{\%r10} for this purpose.
When a caller wishes to make a
function call, it first stores \texttt{\%r10} (which holds the address of
the function to which the caller itself should return) to the shadow stack
and then loads the return address for the \emph{callee} into \texttt{\%r10}.
On function return, control is redirected to the address
saved in \texttt{\%r10} via a jump-through-register instruction.
We could also, of course, compare the main and shadow stack copies of
the return address and raise an exception if they differ, but there
seems to be no point: when running IskiOS, the main-stack return address
is entirely unused, and attackers have no motivation to change it.

After each call, the compiler inserts code to load
the return address from the shadow stack back into \texttt{\%r10}.  Since the
shadow stack is readable, no \texttt{wrpkru} instruction is needed for
this reload.
By writing to the shadow stack only when making a
nested call, \System\ avoids the expense of a \texttt{wrpkru}
instruction when a function calls no other function---i.e., when it is
a dynamic leaf function.  Furthermore, when a function makes more
than one nested call, the expense is incurred only once---more on this
in Section~\ref{section:opts:swo}.

\textbf{Interrupt Handling}
One limitation of our shadow stack design is that it does not leverage
the hardware's return address predictor, which is primed by
\texttt{call} instructions and employed by \texttt{ret}.  It is tempting
to pursue a convention that
points the stack pointer into the shadow stack and uses a real
\texttt{ret} instruction.
Unfortunately, such an approach is incompatible
with the processor's interrupt dispatch mechanism:
a hardware interrupt that occurs when the stack pointer is pointing to the
write-protected shadow stack will cause the kernel to crash when it attempts
to save the processor state on what the kernel thinks is the kernel stack
before executing the interrupt handler.

For two reasons, {\System} must save and restore the
{\tt pkru} on the shadow stack during system call, trap, and
interrupt dispatch.  First,
application code can change that register~\cite{IntelArchManual21},
so {\System} must preserve the value needed in user code.
Second, interrupts can occur while an OS kernel function has enabled
write access to the shadow stack;
{\System} must disable write access to the shadow stack
prior to executing the interrupt handler and restore the \texttt{pkru}
when returning from the interrupt.

\section{Performance Optimizations}
\label{section:opts}

Hedayati et al.~\cite{Hodor:ATC19} report and our experiments confirm that the
cost of executing a \texttt{wrpkru} instruction is roughly 26~cycles. This can
lead
to overheads ranging from negligible to significant, depending on how
protection keys are used.  For
example, enabling XOM requires a single \texttt{pkru} update to drop the
access permissions for code---after that, it's free.
Protected shadow stacks, on the other hand, require {\System} to execute two
\texttt{wrpkru} instructions every time it writes a return address to the
shadow stack: one to enable access and one to
disable it again, for a cost of about $52$~additional cycles in each
dynamic call.
To avoid this cost whenever possible, we explore two optimizations to
our shadow stack.

\subsection{Shadow Write Optimization}
\label{section:opts:swo}

We observe that {\System} can avoid writing the return value to the
shadow stack if the desired value is already there.
In particular, if a function calls 20 other functions, it only has to
save its return address when it makes the first of these calls---the
rest will be \texttt{wrpkru}-free.
Likewise, if a recursive function has been optimized using tail-call
optimization, its caller will use a \texttt{jmp}
instruction instead of a \texttt{call}.
In this case, the return address has already been pushed to the
shadow stack; there is no need to write it again.

To leverage this observation, we designed the
\emph{Shadow Write Optimization (SWO)}.  With SWO, {\System} adds code
to every callsite that first checks whether the word in the shadow
stack to which the return address will be written already contains the
return address.  If it does, the word is not written
a second time.  Experiments in Section~\ref{section:eval} confirm that the
reduction in dynamic executions of \texttt{wrpkru} leads to a
dramatic overall improvement in performance.

\subsection{Aggressive Inlining}
\label{section:opts:inline}
A more obvious optimization opportunity is to minimize the number of function
calls, which in turn eliminates the need to save the return address on
the shadow stack. To achieve this, we aggressively increased the
inlining threshold option to the compiler until we
saw no additional performance improvement. Experiments in
Section~\ref{section:eval} confirm that aggressive inlining almost always leads
to better performance at the cost of increased code size.

\section{Implementation}
\label{section:impl}

We implemented {\System} as a set of patches
to the Linux v5.7 kernel and the Clang/LLVM~\cite{LLVM:CGO04} 10.0.0 compiler.

\subsection{Kernel Modifications}
\label{section:impl:kernel}

In our {\System} prototype, we did not implement the debug watchpoint
arming described in
Section~\ref{section:pkk:safety}. We did, however, use Hodor's
code scanner~\cite{Hodor:ATC19} to confirm the absence of {\tt wrpkru} instructions in the
binaries used in Section~\ref{section:eval:macrobenchmarks}.
We also did not implement the changes to physmap described in
Sections~\ref{section:usecases:xom} and~\ref{section:usecases:shadow},
but we do not expect these to impact performance.
Finally, we carefully considered the PKU vulnerabilities described by Connor
et al.~\cite{PKUPitfalls:SEC20}.  We concluded that, while the design
of {\System} is resistant to these attacks, our prototype
implementation would require additional engineering to resist
certain attacks on memory permissions
(e.g., using \texttt{process\_vm\_readv()}~\cite{process-vm-readv} to read data
from a process's own address space, ignoring the PKU permissions).
Fixing the {\System} prototype to make it complete (and therefore
resistant to these attacks) requires that we add simple checks to the
relevant memory-related system calls.  The changes are straightforward,
and we expect their performance impact to be negligible.
We built {\System} as three separate
patches to the Linux kernel, adding support for
protection keys, execute-only memory, and shadow stacks, respectively.
We describe each patch below.

\paragraph{{\System}-PKK} This patch enables PKU for all of
virtual memory. We mark all pages as user mode by setting the
U/S bit in every PTE\@. Since our design associates
PKEYs 0--7 with kernel space and PKEYs 8--15 with user space,
we change the default protection key for user pages from 0 to 8.
By default, the kernel can access only pages with PKEY 0
(and PKEYs 8--15 when it wants to access user memory);
user processes can access only pages with PKEY 8.

We also modified Linux to save and restore the \texttt{pkru}
on kernel entry and exit.  As an optimization, we arranged for interrupt
dispatch code to elide the {\tt wrpkru} instruction if the \texttt{pkru}
already holds the correct value; this improves performance if a trap or
interrupt occurs while the OS kernel is running.
We disable this optimization in the shadow stack patch, as
{\System} must change {\tt pkru} on every interrupt, trap, and system call.

\paragraph{{\System}-XOM} This patch changes the 4-bit protection key in every
PTE that maps a kernel code page to the value 1.  Read and write access to
pages with PKEY~1 is already disabled in the \texttt{pkru}.

\paragraph{{\System}-SS} This patch (1) doubles (from 16 to 32\,KB) the size
of every stack in the kernel (i.e., every per-thread stack, per-cpu
interrupt stack, etc.) and uses the upper half as a shadow stack, with
write access disabled using PKK\@;
(2) modifies kernel entry/exit code to save/restore the
{\tt pkru} register to/from the shadow stack, and
(3) updates assembly code to conform to the new calling convention
(Sec.~\ref{section:usecases:shadow}).
After updating the standard {\tt CALL\_NOSPEC} macro, we had
to manually modify fewer than 100 additional call sites.

\subsection{Compiler Instrumentation}
\label{section:impl:llvm}

We implemented two separate plugins to the
LLVM 10.0.0 compiler~\cite{LLVM:CGO04} to implement our prototype.
Both extend the LLVM code generator with \emph{MachineFunction} passes:
one adds control-flow assertions on indirect branches and return instructions
to enable \texttt{SMEP} functionality; the other implements shadow stacks.

\subsubsection{SMEP Implementation}
\label{section:impl:llvm:smep}
\begingroup

\begin{figure}[tb]
\lstset{
 language=[x86masm]Assembler,  
 morekeywords={CDQE,CQO,CMPL,CMPB,CMPW,CMPSQ,CMPXCHG16B,JRCXZ,JMPQ,ADDQ,LODSQ,MOVSXD,%
                  POPFQ,PUSHFQ,SCASQ,STOSQ,IRETQ,RDTSCP,SWAPGS,TESTB,UD2,LEAQ,
                  CALLQ,BT,JB, %
                  VMFUNC,WRPKRU,XORL,RDPKRU,MOVL,MOVQ,LFENCE,RETQ, %
                  rax,rdx,rcx,rbx,rsi,rdi,rsp,rbp,rip, %
                  r8,r8d,r8w,r8b,r9,r9d,r9w,r9b, %
                  r10,r10d,r10w,r10b,r11,r11d,r11w,r11b, %
                  r12,r12d,r12w,r12b,r13,r13d,r13w,r13b, %
                  r14,r14d,r14w,r14b,r15,r15d,r15w,r15b},
  belowcaptionskip=0.8\baselineskip,
  xleftmargin=2em,
  basicstyle=\footnotesize\ttfamily,
  commentstyle=\itshape\color{purple!40!black},
  keywordstyle=\color{black}\bfseries,
  numberstyle=\color{codegray},
  numbers=left,
  numbersep=7pt,
  breaklines=true,
  commentstyle=\color{codegreen},
  escapeinside={{*@}{@*}},
}
\begin{lstlisting}[caption={SMEP: Indirect Branch Instrumentation.}, label={lst:smep-branch}]
; check most significant bit of target address
bt $63, *@\%@*TargetReg		*@\label{bt}@*

; if address is in kernel space, skip exception
jb L1				*@\label{skiptrap}@*

; raise exception
ud2						*@\label{trap}@*

L1:
; indirect call/jmp *TargetReg	*@\label{branch}@*
\end{lstlisting}

\begin{lstlisting}[caption={SMEP: Epilogue Instrumentation.}, label={lst:smep-epilogue}]
; load the return address to *@\color{codegreen}\%@*r10
movq 	[*@\%@*rsp], *@\%@*r10 *@\label{loadr10}@*

; check most significant bit of target address
bt $63, *@\%@*r10		*@\label{checkr10}@*

; if address is in kernel space, skip exception
jb L1				*@\label{skiptrap2}@*

; raise exception
ud2					*@\label{trap2}@*

L1:
; adjust *@\color{codegreen}\%@*rsp
addq 0x8, *@\%@*rsp 		*@\label{adjust-rsp}@*

; return
jmpq **@\%@*r10			*@\label{return}@*
\end{lstlisting}
\vspace{-2ex}
\end{figure}
\endgroup

As suggested in Section~\ref{section:pkk:smep},
Listing~\ref{lst:smep-branch} shows instrumentation
inserted before each indirect branch.
Assuming that \texttt{\%TargetReg} contains the target address, the plugin
prepends the indirect \texttt{call} or \texttt{jmp} (line~\ref{branch})
with a \texttt{bt} (line~\ref{bt}) instruction
that stores the most significant bit of the target address in the \texttt{CF}
flag of the status register.
A \texttt{jb} (line~\ref{skiptrap}) instruction
continues regular execution if the address was in kernel space;
otherwise, an undefined instruction
\texttt{ud2/0x0f0b} (line~\ref{trap}) will raise an invalid opcode exception,
and the kernel will crash.

Our \texttt{SMEP} plugin also instruments function epilogues to ensure
that functions do not return to user space. As
Listing~\ref{lst:smep-epilogue} shows, the plugin loads the target address
to \texttt{\%r10} (line~\ref{loadr10}), performs the same test as before
(lines~\ref{checkr10}--\ref{trap2}),
adjusts the \texttt{\%rsp} (line~\ref{adjust-rsp}) to emulate
a function return, and replaces the \texttt{ret} instruction with a
\texttt{jmp} through \texttt{\%r10} (line~\ref{return}).

\subsubsection{Shadow Stack Implementation}
\label{section:impl:llvm:ss}

Our second compiler plugin transforms every callsite to save the return address
on the shadow stack and every function epilogue to use the return
value on the shadow stack as
Section~\ref{section:usecases:shadow} describes.  To support our new calling
convention, we modified the compiler to reserve
register \texttt{\%r10}.

\begingroup
\begin{figure}[p]
\lstset{
 language=[x86masm]Assembler,  
 morekeywords={CDQE,CQO,CMPL,CMPB,CMPW,CMPSQ,CMPXCHG16B,JRCXZ,JMPQ,ADDQ,LODSQ,MOVSXD,%
                  POPFQ,PUSHFQ,SCASQ,STOSQ,IRETQ,RDTSCP,SWAPGS,TESTB,UD2,LEAQ,
                  CALLQ,BT,JB, %
                  VMFUNC,WRPKRU,XORL,RDPKRU,MOVL,MOVQ,LFENCE,RETQ, %
                  rax,rdx,rcx,rbx,rsi,rdi,rsp,rbp,rip, %
                  r8,r8d,r8w,r8b,r9,r9d,r9w,r9b, %
                  r10,r10d,r10w,r10b,r11,r11d,r11w,r11b, %
                  r12,r12d,r12w,r12b,r13,r13d,r13w,r13b, %
                  r14,r14d,r14w,r14b,r15,r15d,r15w,r15b},
  belowcaptionskip=0.8\baselineskip,
  xleftmargin=2em,
  basicstyle=\footnotesize\ttfamily,
  commentstyle=\itshape\color{purple!40!black},
  keywordstyle=\color{black}\bfseries,
  numberstyle=\color{codegray},
  numbers=left,
  numbersep=7pt,
  breaklines=true,
  commentstyle=\color{codegreen},
  escapeinside={{*@}{@*}},
}
\begin{lstlisting}[caption={Shadow Stack Callsite Instrumentation.},
label={lst:ss-callsite}]
  | ; skip shadow write if address already present
  |  cmp  *@\%@*r10, [*@\%@*rsp-4*PGSIZE] *@\label{swo-cmp}@*
  |  je   Lcall	*@\label{skip-sw}@*

; spill *@\color{codegreen}\%@*rax, *@\color{codegreen}\%@*rcx, *@\color{codegreen}\%@*rdx *@\label{spill-regs}@*

; flip shadow stack pkru bit to enable write access
xor rcx, rcx *@\label{clear-rcx}@*
rdpkru		 *@\label{rdpkru}@*
xorl	0x8, *@\%@*eax *@\label{toggle-bit1}@*
wrpkru 		 *@\label{wrpkru1}@*

; ensure execution is in kernel mode
movq 	*@\%@*cs, *@\%@*rcx *@\label{load-cs}@*
testb	$3, *@\%@*cl   *@\label{byte-test}@*
je   	Lskip1 	  *@\label{skip1}@*
Ltrap1:	*@\label{Ltrap1}@*
ud2  *@\label{ud1}@*

Lskip1:	*@\label{Lskip1}@*
; ensure access to xom has not been enabled
cmpb 0xc, *@\%@*ax *@\label{hodor-test}@*
jne Ltrap1 *@\label{trap1}@*

; copy ret. addr. to shadow stack
movq	*@\%@*r10, [*@\%@*rsp-4*PGSIZE]*@\label{shadow-write}@*

; flip shadow stack pkru bit to disable write access
xorl	0x8, *@\%@*eax *@\label{toggle-bit2}@*
wrpkru 		*@\label{wrpkru2}@*

; ensure execution is in kernel mode
movq 	*@\%@*cs, *@\%@*rcx *@\label{load-cs2}@*
testb 	$3, *@\%@*cl *@\label{byte-test2}@*
je  	Lskip2 	*@\label{skip2}@*
Ltrap2:	*@\label{Ltrap2}@*
ud2  *@\label{ud2}@*

Lskip2: *@\label{Lskip2}@*
; ensure access to xom has not been enabled
cmpb 0xc, *@\%@*ax *@\label{hodor-test2}@*
jne Ltrap2 *@\label{jneLtrap2}@*

; restore spilled regs 	*@\label{restore-regs}@*

Lcall:
; copy new ret. addr. to *@\color{codegreen}\%@*r10
leaq	Lret(*@\%@*rip), *@\%@*r10 *@\label{load-r10}@*

; actual callsite
callq	**@\%@*rbx *@\label{call}@*
Lret:

; restore old ret. addr. to *@\color{codegreen}\%@*r10
movq	[*@\%@*rsp-4*PGSIZE], *@\%@*r10 *@\label{restore-r10}@*
\end{lstlisting}
\medskip
\begin{lstlisting}[caption={Shadow Stack Epilogue Instrumentation.}, label={lst:ss-epilogue}]
; adjust *@\color{codegreen}\%@*rsp
addq 0x8, *@\%@*rsp 		*@\label{adjust-ss-rsp}@*

; return
jmpq **@\%@*r10			*@\label{ss-return}@*
\end{lstlisting}
\vspace{-2ex}
\end{figure}
\endgroup

Listing~\ref{lst:ss-callsite} shows {\System}'s callsite instrumentation with
and without \texttt{SWO}.
When the shadow-write optimization is disabled,
our compiler pass enables access to the shadow stack at each callsite by
using \texttt{wrpkru} to clear the write-disable (\texttt{WD}) bit for
PKEY~3 (lines~\ref{clear-rcx}--\ref{wrpkru1}).
It then copies the return address of the currently executing function
to the shadow stack (line~\ref{shadow-write})
and executes another \texttt{wrpkru} (line~\ref{wrpkru2})
to disable write access again.  Note that since \texttt{rdpkru}
and \texttt{wrpkru} force us to zero \texttt{\%ecx} and \texttt{\%edx},
the compiler must spill (line~\ref{spill-regs})
and restore (line~\ref{restore-regs}) these registers.
Before the call instruction (line~\ref{call}), our compiler
loads the callee's return address into \texttt{\%r10} (line~\ref{load-r10}).
(The return address saved to the stack by the \texttt{call} instruction
is never used.)
After the call site, the compiler restores the return address in \texttt{\%r10}
from the shadow stack (line~\ref{restore-r10}).
When \texttt{SWO} is enabled, the compiler adds a simple check
(lines~\ref{swo-cmp}--\ref{skip-sw})
which ensures that the return address is pushed
onto the shadow stack only if the value already there differs.

Listing~\ref{lst:ss-epilogue} shows {\System}'s instrumentation of function
epilogues. The epilogue code
adjusts the stack pointer (line~\ref{adjust-ss-rsp}) to discard the return
address on the main stack and performs an indirect jump
to the address in \texttt{\%r10} (line~\ref{ss-return}).

\section{Evaluation}
\label{section:eval}

We evaluated the performance overhead that {\System} incurs for its
\texttt{PKK}-based user/kernel isolation mechanism,
execute-only memory, and protected shadow stacks.
We also evaluated the shadow-write and inlining optimizations
described in Section~\ref{section:opts}.
Finally, we evaluated the impact of {\System} on the size of the kernel
code segment.
We report numbers for the following configurations:

\begin{itemize}
  \item{\textbf{vanilla}:}
  Unmodified Linux kernel v5.7

  \item{\textbf{pkk}:}
  {\System} kernel with protection keys for kernel space

  \item{\textbf{pkk+xom}:}
  {\System} with \texttt{PKK} and execute-only memory

  \item{\textbf{pkk+ss}:}
  {\System} with \texttt{PKK} and unoptimized shadow stacks

  \item{\textbf{pkk+ss-swo}:}
  {\System} with \texttt{PKK}, shadow stacks, and the shadow-write
  optimization (\texttt{SWO})

  \item{\textbf{pkk+ss-swo-inline}:}
  {\System} with \texttt{PKK}, shadow stacks with \texttt{SWO},
  and an increased inlining threshold

  \item{\textbf{pkk+xom+ss-swo-inline}:}
  {\System} with \texttt{PKK}, \texttt{XOM},
  shadow stacks with \texttt{SWO}, and an increased inlining threshold
\end{itemize}

In all the above configurations, we disabled Kernel Address Space
Layout Randomization (KASLR)~\cite{KASLR:LKML}, Kernel Page Table Isolation
(KPTI)~\cite{KPTI:LKML}, and hardware SMAP and SMEP.
We used the LMBench suite~\cite{lmbench} for microbenchmarking and the
Phoronix Test Suite (PTS)~\cite{benchmark:PTS} to measure the performance
impact on real-world applications.
We performed our experiments on a Fedora 32 system equipped with two
3.00\,GHz Intel Xeon Silver~4114 (Skylake) CPUs (2$\times$10 cores, 40
total threads), 16\,GB of RAM and a 1\,TB Seagate 7200 RPM disk.
For all our tests, we loaded the \texttt{intel\_pstate} performance scaling
driver into the kernel to prevent the processor from reducing frequency (for
power saving) during our experiments.
For the networking experiments, we ran client and server processes on the same
machine. We used the default settings for all benchmarks.

\subsection{Microbenchmarks}
\label{section:eval:microbenchmarks}

We used LMBench~\cite{lmbench} v.3.0-a9 to measure the latency and bandwidth
overheads imposed by {\System} on basic kernel operations. In particular,
our set of microbenchmarks measures the latency of critical I/O
system calls (\texttt{open()/\linebreak[1]close()}, \texttt{read()/\linebreak[1]write()},
\texttt{select()}, \texttt{fstat()}, \texttt{stat()},
\texttt{mmap()/\linebreak[1]munmap()}),
execution mode switches (\texttt{null} system call), and
inter-process context switches.
We also measured the impact on process creation followed by
\texttt{exit()}, \texttt{execve()}, and \texttt{/bin/sh},
as well as the latency of
signal installation (via \texttt{sigaction()}) and delivery,
protection faults, and page faults.
Finally, we measured the latency overhead on pipe and socket
I/O (TCP, UDP, and Unix domain sockets) and the bandwidth
degradation on pipes, sockets, and
file and memory-mapped I/O\@. We report the arithmetic mean of
10 runs for each microbenchmark.

\subsubsection{{\System}'s Performance}
\label{section:eval:microbenchmarks:iskios}

\begin{figure*}
  \centering
\begin{subfigure}{0.8\textwidth}
  \caption{{\System} execution time, normalized to {\tt vanilla} Linux.}
  \label{fig:lmbench:latency}
  \includegraphics[width=\textwidth]{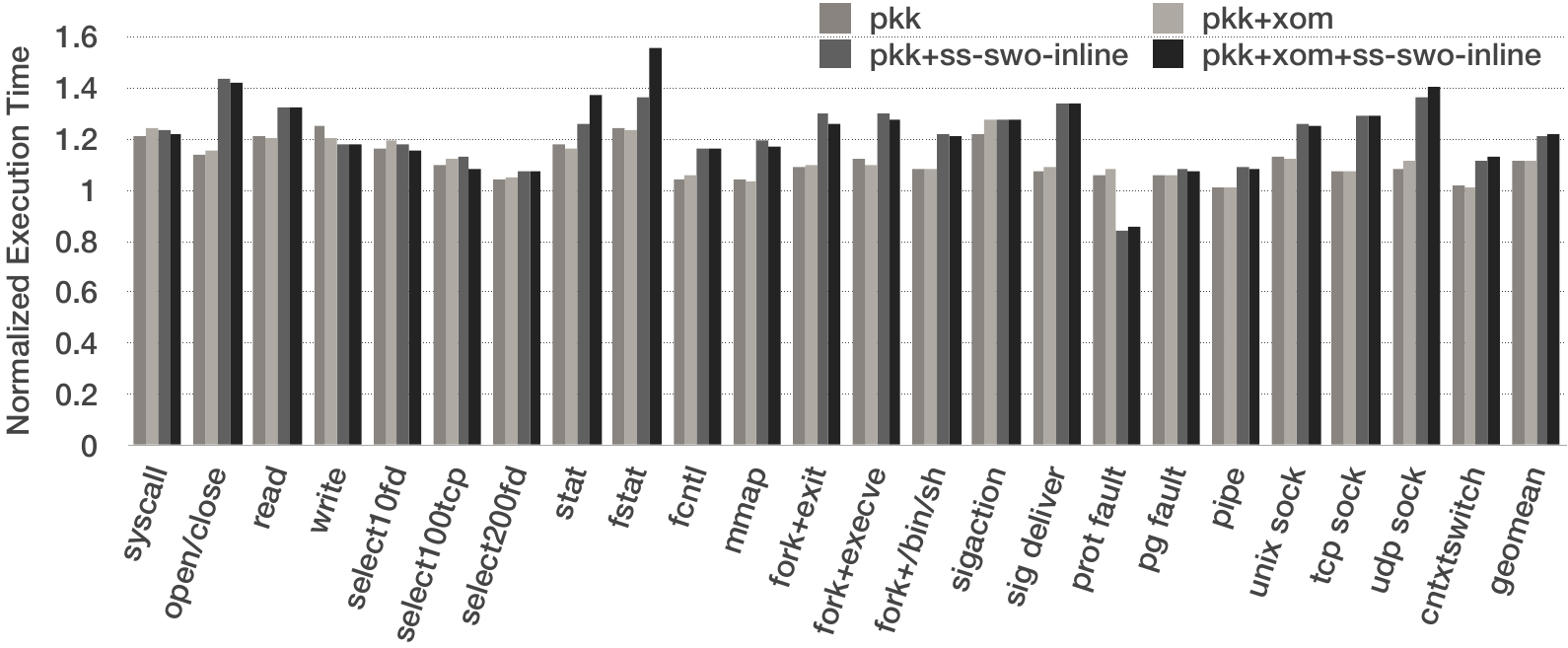}
\end{subfigure}

\vspace{2ex}
\begin{subfigure}{0.8\textwidth}
  \subcaption{Performance impact of shadow stack optimizations.}
  \label{fig:ss-eval:latency}
  \includegraphics[width=\textwidth]{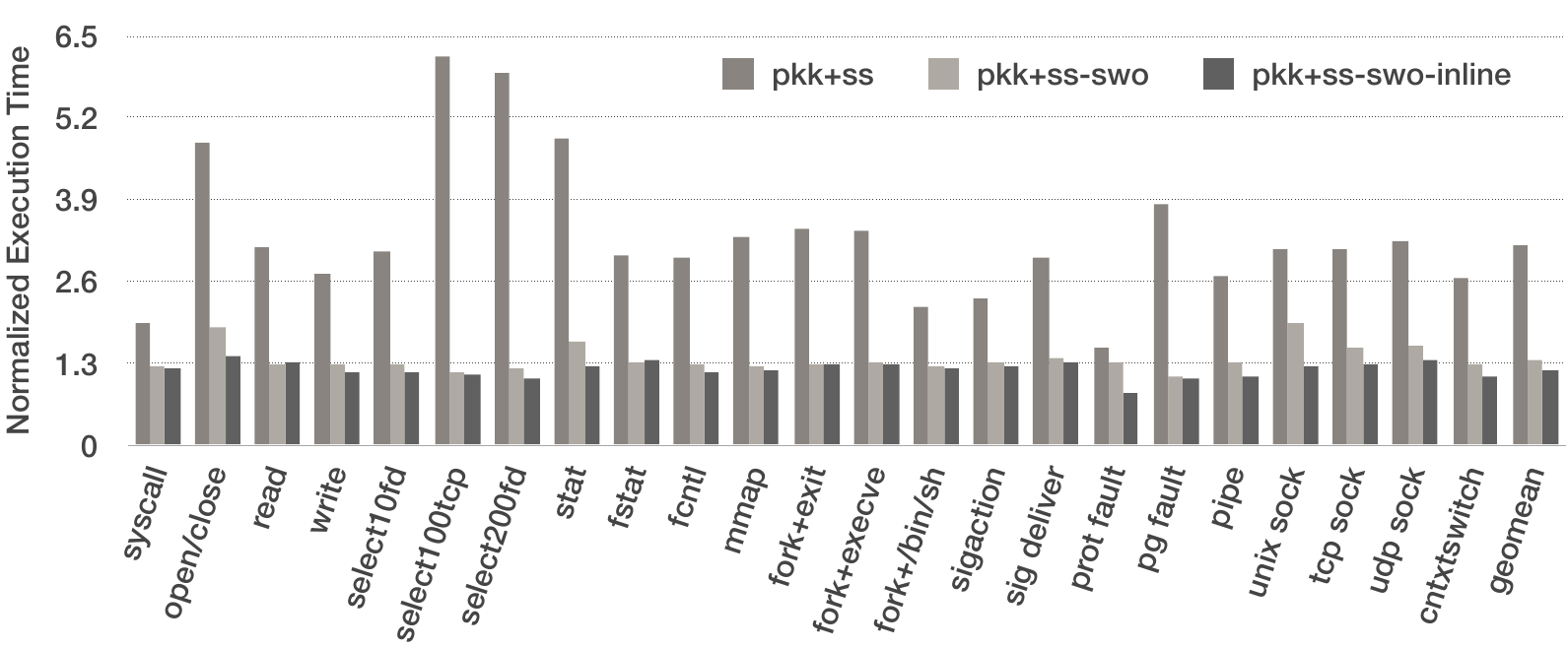}
\end{subfigure}

\vspace{3ex}
\caption{Evaluation of {\System}'s performance on latency microbenchmarks of
the LMBench suite.}
\label{fig:lmbench-results:latency}
\end{figure*}

\begin{figure*}[!hbt]
\hfill
\begin{subfigure}{\columnwidth}
 \subcaption{\raggedright {\System} bandwidth, normalized to {\tt vanilla} Linux.}
  \label{fig:lmbench:bw}
  \includegraphics[width=\textwidth]{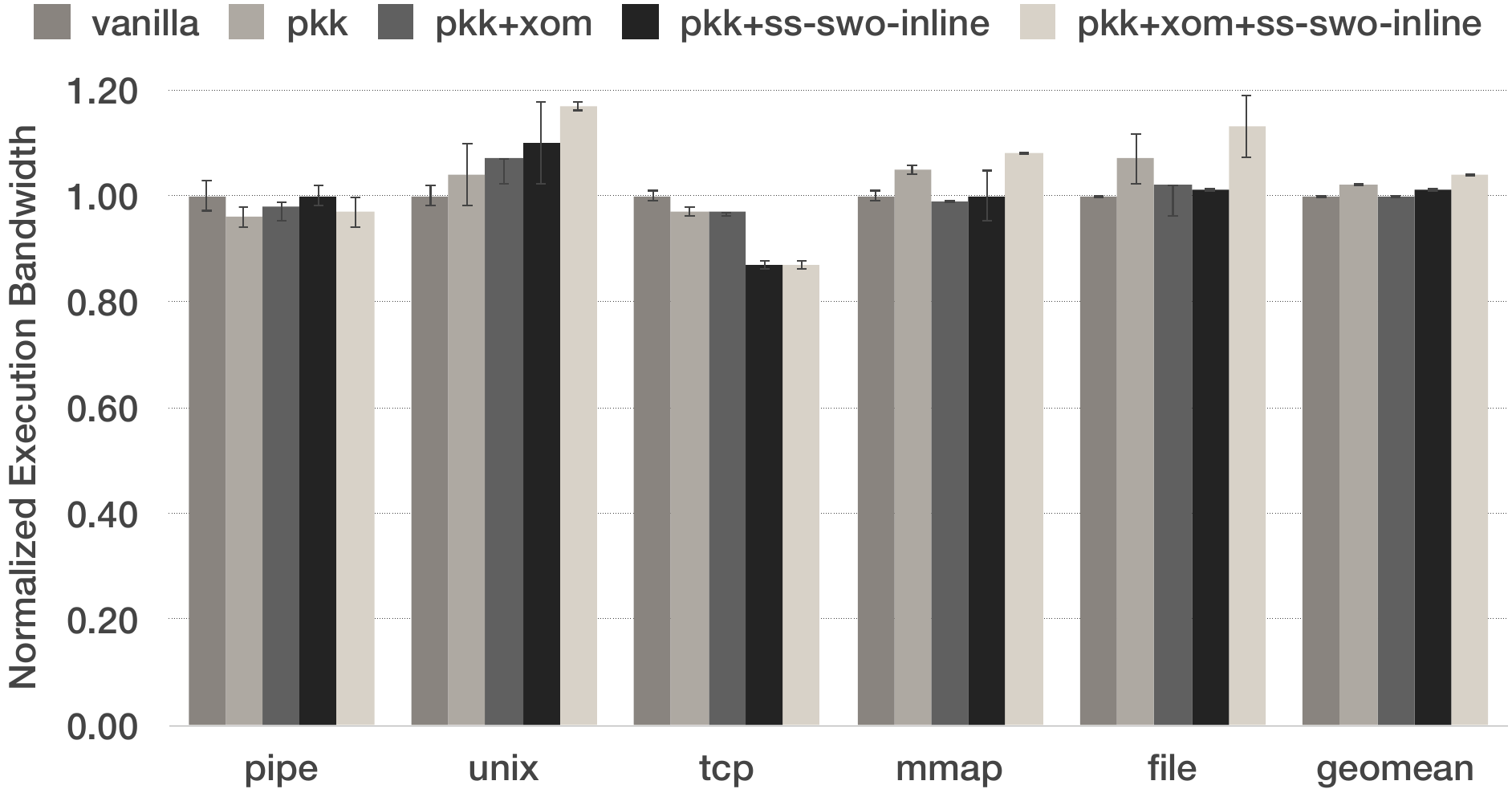}
\end{subfigure}
\hfill
\begin{subfigure}{\columnwidth}
  \subcaption{\raggedright Bandwidth impact of shadow stack optimizations.}
  \label{fig:ss-eval:bw}
  \includegraphics[width=\textwidth]{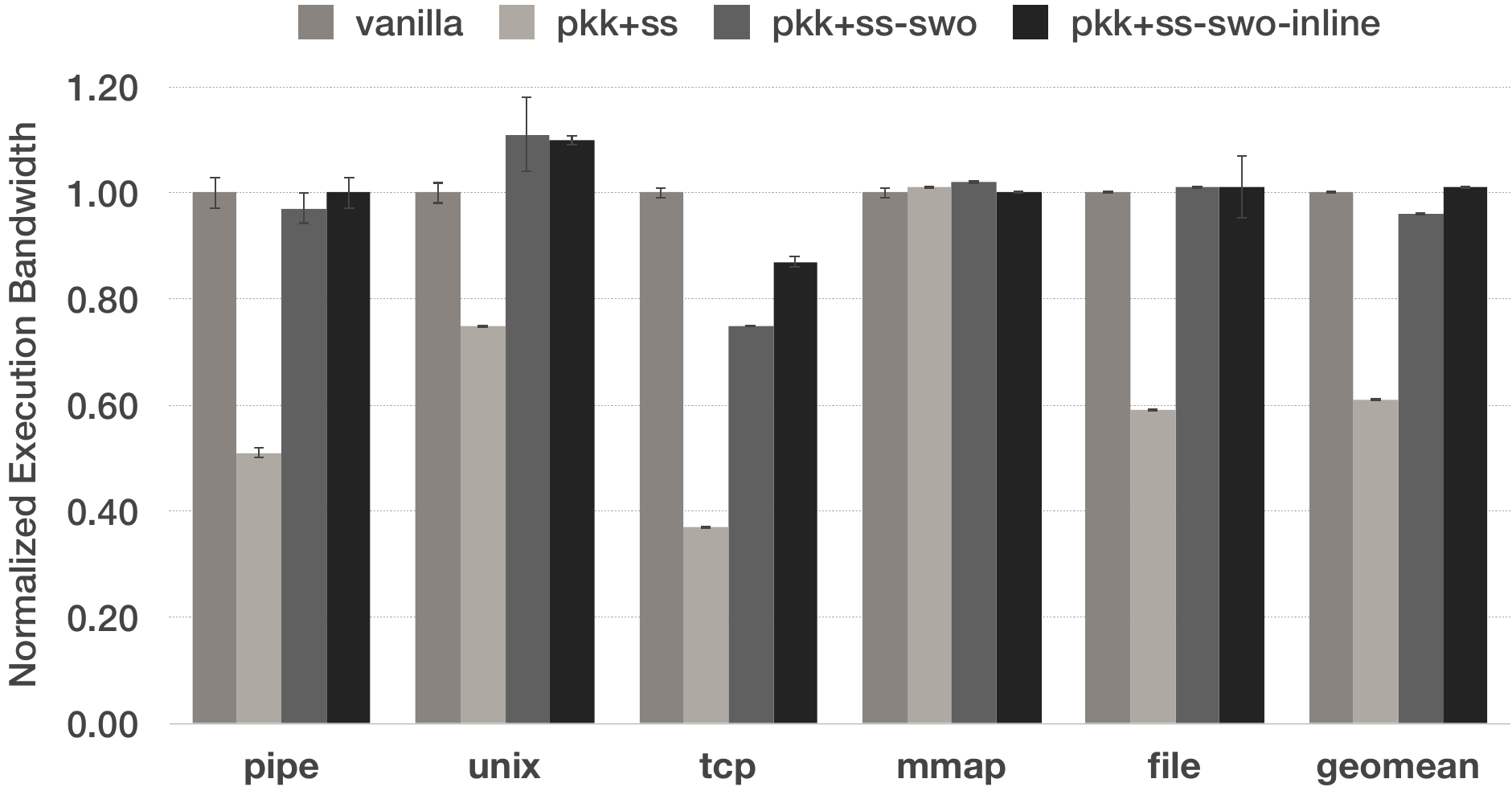}
\end{subfigure}
\hfill\strut
\caption{Evaluation of {\System}'s performance on bandwidth microbenchmarks
of the LMBench suite. The error bars represent the normalized standard
deviation (SD) of 10 runs for each benchmark.}
\label{fig:lmbench-results:bw}
\end{figure*}

Figures~\ref{fig:lmbench-results:latency} and~\ref{fig:lmbench-results:bw}
present our results for latency and bandwidth microbenchmarks,
respectively (absolute performance numbers and standard deviations
appear in Tables~\ref{table:eval:lmbench-lat}
and~\ref{table:eval:lmbench-bw} in Appendix~\ref{section:appendix-a}).
First we discuss {\System}'s effect on the latency of basic kernel operations.
Figure~\ref{fig:lmbench:latency} shows that baseline support for \texttt{PKK}
incurs low overhead for most
microbenchmarks relative to \texttt{vanilla} Linux:
roughly $11\%$ geometric
mean and a maximum of $\sim$25\% for the \texttt{write}
system call.  \texttt{PKK} incurs higher overheads on lightweight system
call tests.
For example, the \texttt{null} test calls \texttt{getppid()}, a system call
that does minimal work inside the kernel, in a loop to estimate the
round-trip latency of entering and exiting the kernel. The
\texttt{read} test reads a single byte from \texttt{/dev/zero}, and
the \texttt{write} test writes one byte to \texttt{/dev/null}, estimating a
lower bound on the latency of any user/kernel interaction.  \texttt{PKK} adds
less than ten instructions to the OS kernel
entry/exit path and two instructions for every indirect branch in kernel code.
As expected, the marginal impact on kernel operations (except for
extremely small services as described above) is negligible.

{\System}'s \texttt{XOM} incurs virtually no overhead (when accounting
for standard error) compared to \texttt{PKK}, and an average (geomean)
of $\sim$12\% for latency microbenchmarks compared to
\texttt{vanilla} Linux.
The \texttt{pkk+ss-swo-inline} kernel (fully optimized shadow stack,
with \texttt{SWO} and aggressive inlining)
incurs $\sim$21\% overhead compared to \texttt{vanilla}
and $10$\% compared to \texttt{PKK}.
Finally, the \texttt{pkk+xom+ss-swo-}\linebreak[1]\texttt{inline} kernel,
with both \texttt{XOM} and a fully-optimized shadow stack,
incurs $\sim$22\% overhead compared to vanilla Linux.

Figure~\ref{fig:lmbench:bw} shows the effect of {\System} on
the bandwidth of various inter-process communication mechanisms.
{\System} imposes the most overhead---roughly $16\%$ over
\texttt{vanilla}---when moving data over a TCP/IP socket with
both \texttt{XOM} and a fully-optimized shadow stack enabled. The
complexity of the underlying TCP/IP protocol combined with the speed
of network communication to localhost explains this result:
TCP performs congestion, flow, and transmission-error control, and the
network stack performs many function calls both within and between layers.
These calls
increase the number of instructions
executed (including the relatively expensive \texttt{wrpkru}) when
the shadow stack is enabled, and there is no latency from a real
network to hide the additional overhead.

\subsubsection{PKK Overhead Breakdown}
\label{section:eval:microbenchmarks:pkk}

To assess the contribution to \texttt{PKK} overhead of our
software-based \texttt{SMEP} support, we disabled the LLVM \texttt{SMEP} pass
and reevaluated \texttt{PKK}'s performance on the LMBench microbenchmarks.
Due to space, we summarize the results;
Appendix~\ref{section:appendix-a} contains full details for
interested readers.
Our results show that our SMEP instrumentation causes most of PKK's overhead:
disabling SMEP reduces the geometric mean PKK overhead from
11\% to 5\% on the LMBench latency benchmarks and from
2\% to 1\% on the LMBench bandwidth benchmarks.
We believe this overhead is due to the extra instructions added to
every indirect branch and the loss of utilizing the processor's
return predictor.

\subsubsection{Shadow Stack Optimizations}
\label{section:eval:microbenchmarks:opts}

Figures~\ref{fig:ss-eval:latency} and~\ref{fig:ss-eval:bw} assess the
importance of our shadow-write and inlining optimizations for the
\System\ shadow stack.
The unoptimized shadow stack implementation (\texttt{pkk+ss})
adds an \texttt{rdpkru} and two \texttt{wrpkru} instructions
to every callsite in a function, incurring a maximum
overhead of nearly $520\%$ for the \texttt{select100tcp} latency
microbenchmark. Across all latency microbenchmarks, the unoptimized shadow
stack incurs a geometric mean overhead of $\sim$217\% compared to our
baseline. These results are consistent with the recent work of Burow
et. al~\cite{SoKShiningLight:SP19}, who evaluate
various ISA features, including Intel PKU, for protecting a
user-space shadow stack.  In particular, Burow et. al report a maximum
overhead of $\sim$420\% on the \texttt{SPEC CPU2006}~\cite{SPEC:CPU2006}
benchmark suite when using Intel PKU to write-protect the shadow stack.

When shadow-write optimization (\texttt{SWO}) is enabled, however (i.e.,
in the \texttt{pkk+ss-swo} kernel), overheads decrease
dramatically, to a maximum of $\sim$97\%  and a geometric mean of $\sim$37\%.
When both \texttt{SWO} and the inlining optimization are enabled
(\texttt{pkk+ss-swo-inline}), the overheads drop to a maximum
of $\sim$43\% and a geomean of $\sim$21\%.
While differences in control flow and instruction mix make it difficult to
directly compare user and kernel workloads, our results nonetheless
demonstrate the viability of PKU for intra-address-space
isolation, even for mechanisms like shadow stacks, which require
frequent domain switching.
In particular, it should be possible to significantly improve the
performance of user-space shadow stacks~\cite{SoKShiningLight:SP19}
using SWO and (perhaps) more aggressive inlining.

\subsection{Macrobenchmarks}
\label{section:eval:macrobenchmarks}
\begin{table*}[tb]
\caption{{\System} run-time overhead ($\%$ over {\tt vanilla} Linux) on
  applications from the Phoronix Test Suite.  The final row reports
  average change in the reported metric (geometric mean).}
\label{table:eval:phoronix}
\setlength{\tabcolsep}{3pt} 
\centering
\begin{LARGE}
\resizebox{\textwidth}{!}{%
\begin{tabular}{ |l|rr|r|r|r|r|r|r|r|r|}
 \hline
                            \multirow{2}{*}{\bf Benchmark}
  & \multicolumn{2}{c|}{\multirow{2}{*}{\bf vanilla}}
  & \multicolumn{2}{c|}{\bf pkk}
  & \multicolumn{2}{c|}{\bf pkk+xom}
  & \multicolumn{2}{c|}{\bf pkk+ss-swo-inline}
  & \multicolumn{2}{c|}{\bf pkk+xom+ss-swo-inline}\\
  \cline{4-11}
  & & &Mean $\pm$ RSD &Overhead &Mean $\pm$ RSD &Overhead &Mean $\pm$ RSD &Overhead &Mean $\pm$ RSD &Overhead \\
  \hline
  {\bf Apache}          
    & $24626 \pm \phantom{0}2.20\%$ & $\mathit{req/s}$
    & $20531 \pm \phantom{0}1.30\%$ & $16.63\%$ 
    & $21294 \pm \phantom{0}1.81\%$ & $13.53\%$
    & $16048 \pm \phantom{0}0.95\%$ & $34.83\%$
    & $15141 \pm \phantom{0}2.40\%$ & $38.51\%$\\
  {\bf Build Apache}
    & $33.24 \pm \phantom{0}1.39\%$ & $\mathit{sec}$
    & $33.38 \pm \phantom{0}0.42\%$ & $0.43\%$
    & $33.58 \pm \phantom{0}0.22\%$ & $1.04\%$
    & $33.82 \pm \phantom{0}0.56\%$ & $1.75\%$
    & $34.08 \pm \phantom{0}0.45\%$ & $2.54\%$\\
  {\bf Build PHP}
    & $71.00 \pm \phantom{0}0.47\%$ & $\mathit{sec}$
    & $71.95 \pm \phantom{0}1.68\%$ & $1.34\%$
    & $71.83 \pm \phantom{0}0.56\%$ & $1.17\%$
    & $72.34 \pm \phantom{0}0.27\%$ & $1.90\%$
    & $72.41 \pm \phantom{0}0.47\%$ & $1.99\%$ \\
  {\bf C-Ray}
    & $148.30 \pm 46.61\%$ & $\mathit{sec}$
    & $117.72 \pm 28.45\%$ & $-20.62\%$
    & $114.85 \pm 43.71\%$ & $-22.56\%$
    & $155.53 \pm 22.89\%$ & $4.88\%$
    & $181.50 \pm 27.93\%$ & $22.39\%$ \\
  {\bf CompileBench}
    & $296.84 \pm \phantom{0}6.59\%$ & $\mathit{MB/s}$
    & $247.43 \pm \phantom{0}3.26\%$ & $16.65\%$
    & $258.08 \pm \phantom{0}1.23\%$ & $13.06\%$
    & $219.27 \pm \phantom{0}4.19\%$ & $26.13\%$
    & $215.28 \pm \phantom{0}3.43\%$ & $27.48\%$\\
  {\bf Crafty}
    & $5038443 \pm \phantom{0}0.25\%$ & $\mathit{nodes/s}$
    & $5016383 \pm \phantom{0}0.17\%$ & $0.44\%$
    & $5031947 \pm \phantom{0}0.06\%$ & $0.13\%$
    & $5042075 \pm \phantom{0}0.04\%$ & $-0.07\%$
    & $5035883 \pm \phantom{0}0.09\%$ & $0.05\%$\\
  {\bf Cryptsetup-sha512}
    & $1016062 \pm \phantom{0}0.00\%$ & $\mathit{iter/s}$
    & $1010211 \pm \phantom{0}0.58\%$ & $0.58\%$
    & $1014096 \pm \phantom{0}0.00\%$ & $0.19\%$
    & $1012801 \pm \phantom{0}0.40\%$ & $0.32\%$
    & $1008968 \pm \phantom{0}1.05\%$ & $0.70\%$\\
  {\bf Cryptsetup-whirlpool}
    & $426949 \pm \phantom{0}0.43\%$ & $\mathit{iter/s}$
    & $424871 \pm \phantom{0}0.32\%$ & $0.49\%$
    & $425561 \pm \phantom{0}0.32\%$ & $0.33\%$
    & $426020 \pm \phantom{0}0.19\%$ & $0.22\%$
    & $426944 \pm \phantom{0}0.00\%$ & $0.00\%$\\
  {\bf Ebizzy}
    & $524560 \pm \phantom{0}3.49\%$ & $\mathit{records/s}$
    & $515582 \pm \phantom{0}0.61\%$ & $1.71\%$
    & $531721 \pm \phantom{0}4.19\%$ & $-1.37\%$
    & $537828 \pm \phantom{0}4.07\%$ & $-2.53\%$
    & $514557 \pm \phantom{0}2.08\%$ & $1.91\%$\\
  {\bf Encode FLAC}
    & $18.69 \pm \phantom{0}3.30\%$ & $\mathit{sec}$
    & $18.42 \pm \phantom{0}1.08\%$ & $-1.44\%$
    & $18.93 \pm \phantom{0}2.27\%$ & $1.28\%$
    & $19.24 \pm \phantom{0}3.01\%$ & $2.98\%$
    & $18.64 \pm \phantom{0}2.44\%$ & $-0.23\%$\\
  {\bf Encode MP3}
    & $16.95 \pm \phantom{0}0.23\%$ & $\mathit{sec}$
    & $16.95 \pm \phantom{0}0.27\%$ & $0.02\%$
    & $16.96 \pm \phantom{0}0.28\%$ & $0.07\%$
    & $16.95 \pm \phantom{0}0.21\%$ & $0.02\%$
    & $16.95 \pm \phantom{0}0.31\%$ & $0.05\%$ \\
  {\bf FS-Mark}
    & $24.7 \pm \phantom{0}0.84\%$ &$\mathit{files/s}$
    & $24.7 \pm \phantom{0}3.28\%$ & $0.00\%$
    & $24.8 \pm \phantom{0}0.23\%$ & $-0.40\%$
    & $24.3 \pm \phantom{0}0.68\%$ & $1.62\%$
    & $22.5 \pm \phantom{0}0.26\%$ & $8.91\%$\\
  {\bf Fhourstones}
    & $8816.50 \pm \phantom{0}0.02\%$ & $\mathit{Kpos/s}$
    & $8812.30 \pm \phantom{0}0.28\%$ & $0.05\%$
    & $8797.00 \pm \phantom{0}0.33\%$ & $0.22\%$
    & $8834.60 \pm \phantom{0}0.04\%$ & $-0.21\%$
    & $8807.50 \pm \phantom{0}0.08\%$ & $0.10\%$\\
  {\bf Git}
    & $88.74 \pm \phantom{0}0.35\%$ & $\mathit{sec}$
    & $88.87 \pm \phantom{0}0.27\%$ & $0.15\%$
    & $89.03 \pm \phantom{0}0.20\%$ & $0.33\%$
    & $89.37 \pm \phantom{0}0.44\%$ & $0.72\%$
    & $89.94 \pm \phantom{0}0.63\%$ & $1.35\%$\\
  {\bf HMMer}
    & $189.69 \pm \phantom{0}0.19\%$ & $\mathit{sec}$
    & $190.35 \pm \phantom{0}0.04\%$ & $0.35\%$
    & $190.41 \pm \phantom{0}0.13\%$ & $0.38\%$
    & $193.30 \pm \phantom{0}0.14\%$ & $1.90\%$
    & $193.43 \pm \phantom{0}0.15\%$ & $1.97\%$\\
  {\bf Himeno}
    & $2113.80 \pm \phantom{0}0.08\%$ & $\mathit{Mflops}$
    & $2099.09 \pm \phantom{0}0.60\%$ & $0.70\%$
    & $2112.68 \pm \phantom{0}0.16\%$ & $0.05\%$
    & $2112.08 \pm \phantom{0}0.06\%$ & $0.08\%$
    & $2112.72 \pm \phantom{0}0.22\%$ & $0.05\%$\\
  {\bf LuaJIT}
    & $912.22 \pm \phantom{0}0.50\%$ & $\mathit{Mflops}$
    & $912.02 \pm \phantom{0}0.34\%$ & $0.02\%$
    & $904.81 \pm \phantom{0}0.11\%$ & $0.81\%$
    & $915.87 \pm \phantom{0}0.32\%$ & $-0.40\%$
    & $915.56 \pm \phantom{0}0.10\%$ & $-0.37\%$\\
  {\bf NGINX}
    & $26850 \pm \phantom{0}1.96\%$ & $\mathit{req/s}$
    & $22981 \pm \phantom{0}0.52\%$ & $14.41\%$
    & $22646 \pm \phantom{0}1.53\%$ & $15.66\%$
    & $17901 \pm \phantom{0}5.89\%$ & $33.33\%$
    & $16878 \pm \phantom{0}0.20\%$ & $37.14\%$\\
  {\bf OpenSSL}
    & $3788 \pm \phantom{0}0.23\%$ & $\mathit{signs/s}$
    & $3693 \pm \phantom{0}0.45\%$ & $2.42\%$
    & $3792 \pm \phantom{0}0.19\%$ & $-0.11\%$
    & $3792 \pm \phantom{0}0.13\%$ & $-0.13\%$
    & $3791 \pm \phantom{0}0.27\%$ & $-0.08\%$ \\
  {\bf PGBench}
    & $681 \pm \phantom{0}6.64\%$ & $\mathit{trans/s}$
    & $695 \pm \phantom{0}6.90\%$  & $-2.06\%$
    & $679 \pm \phantom{0}8.76\%$  & $0.29\%$
    & $649 \pm \phantom{0}1.68\%$  & $4.70\%$
    & $634 \pm \phantom{0}6.59\%$  & $6.90\%$\\
  {\bf PHPBench}
    & $405719 \pm \phantom{0}0.13\%$ & $\mathit{score}$
    & $405729 \pm \phantom{0}0.10\%$ & $0.01\%$
    & $405094 \pm \phantom{0}0.25\%$ & $0.15\%$
    & $406372 \pm \phantom{0}0.15\%$ & $-0.16\%$
    & $405861 \pm \phantom{0}0.10\%$ & $-0.03\%$\\
  {\bf Parallel BZIP2}
    & $4.36 \pm 16.34\%$ & $\mathit{sec}$
    & $4.36 \pm 16.35\%$ & $0.00\%$
    & $4.37 \pm 16.66\%$ & $0.18\%$
    & $4.24 \pm 18.27\%$ & $-2.75\%$
    & $4.30 \pm 17.66\%$ & $-1.38\%$\\
  {\bf PolyBench-C}
    & $5.03 \pm \phantom{0}4.12\%$ & $\mathit{sec}$
    & $4.91 \pm \phantom{0}4.50\%$ & $-2.25\%$
    & $4.91 \pm \phantom{0}4.88\%$ & $-2.29\%$
    & $4.82 \pm \phantom{0}2.46\%$ & $-4.16\%$
    & $4.83 \pm \phantom{0}2.18\%$ & $-3.90\%$\\
  {\bf PostMark}
    & $4548 \pm \phantom{0}3.10\%$ & $\mathit{trans/s}$
    & $4121 \pm \phantom{0}0.95\%$ & $9.39\%$
    & $4237 \pm \phantom{0}0.00\%$ & $6.84\%$
    & $3607 \pm \phantom{0}2.18\%$ & $20.69\%$
    & $3641 \pm \phantom{0}0.84\%$ & $19.94\%$\\
  {\bf Primesieve}
    & $15.79 \pm \phantom{0}0.87\%$ & $\mathit{sec}$
    & $16.03 \pm \phantom{0}0.65\%$ & $1.53\%$
    & $15.86 \pm \phantom{0}0.20\%$ & $0.46\%$
    & $15.80 \pm \phantom{0}0.44\%$ & $0.08\%$
    & $15.76 \pm \phantom{0}0.26\%$ & $-0.16\%$\\
  {\bf SVT-AV1}
    & $2.95 \pm \phantom{0}1.87\%$  & $\mathit{frames/s}$
    & $2.96 \pm \phantom{0}0.87\%$  & $-0.24\%$
    & $2.92 \pm \phantom{0}1.42\%$  & $1.22\%$
    & $2.95 \pm \phantom{0}1.53\%$  & $-0.10\%$
    & $2.99 \pm \phantom{0}1.99\%$  & $-1.22\%$\\
  {\bf SVT-HEVC}
    & $57.45 \pm 10.37\%$ & $\mathit{frames/s}$
    & $57.05 \pm 10.66\%$ & $0.70\%$
    & $57.04 \pm 10.47\%$ & $0.71\%$
    & $56.65 \pm 10.30\%$ & $1.39\%$
    & $56.70 \pm 10.23\%$ & $1.31\%$\\
  {\bf Stockfish}
    & $27.87M \pm \phantom{0}3.24\%$ & $\mathit{nodes/s}$
    & $28.17M \pm \phantom{0}1.26\%$ & $-1.08\%$
    & $27.11M \pm \phantom{0}2.43\%$ & $2.74\%$
    & $27.59M \pm \phantom{0}2.64\%$ & $1.01\%$
    & $28.12M \pm \phantom{0}3.03\%$ & $-0.91\%$\\
  {\bf Systemd Kernel}
    & $2975 \pm \phantom{0}0.00\%$ & $\mathit{msec}$
    & $2985 \pm \phantom{0}0.00\%$ & $0.34\%$
    & $3065 \pm \phantom{0}0.00\%$ & $3.03\%$
    & $3252 \pm \phantom{0}0.00\%$ & $9.31\%$
    & $3220 \pm \phantom{0}0.00\%$ & $8.24\%$\\
  {\bf TTSIOD 3D Renderer}
    & $319.34 \pm \phantom{0}1.60\%$ & $\mathit{frames/s}$
    & $383.37 \pm \phantom{0}7.65\%$ & $-20.05\%$
    & $377.98 \pm 10.90\%$ & $-18.36\%$
    & $361.43 \pm \phantom{0}8.35\%$ & $-13.18\%$
    & $352.00 \pm \phantom{0}9.29\%$ & $-10.23\%$\\
    \hline
  \multicolumn{3}{|c|}{\bf Geometric Mean Overhead}
        & & 0.39\%
        & & 0.36\%
        & & 3.67\%
        & & 4.87\% \\
  \hline
\end{tabular}
}
\end{LARGE}
\end{table*}

\begin{table*}[bt]
\caption{Macrobenchmarks with highest overhead for \texttt{PKK};
top four system calls for each, with
percent of CPU time spent running in the kernel.} \label{table:eval:strace}
\centering
 \begin{footnotesize}
\begin{tabular}{ |l|c|c|c|c|c|c|c|c|}
 \hline
  {\bf Benchmark}
  &{\bf Syscall} &{\bf \%CPU} &{\bf Syscall} &{\bf \%CPU}
  &{\bf Syscall} &{\bf \%CPU} &{\bf Syscall} &{\bf \%CPU}\\
  \hline
  {\bf CompileBench}    & \texttt{write()}        & $60.57\%$
                        & \texttt{openat()}       & $15.49\%$
                        & \texttt{fstat}          & $10.77\%$
                        & \texttt{close()}        & $5.86\%$ \\

  {\bf Apache - server} & \texttt{futex()}        & $91.00\%$
                        & \texttt{read()}         & \phantom{0}$4.11\%$
                        & \texttt{epoll\_wait()}  & \phantom{0}$1.87\%$
                        & \texttt{accept4()}      & \phantom{0}$0.45\%$ \\

  {\bf Apache - client} & \texttt{epoll\_ctl()}   & $23.29\%$
                        & \texttt{connect()}      & $18.58\%$
                        & \texttt{write()}        & $15.05\%$
                        & \texttt{read()}         & $13.16\%$ \\

  {\bf NGINX - server}  & \texttt{close()}        & $14.61\%$
                        & \texttt{recvfrom()}     & $12.91\%$
                        & \texttt{writev()}       & $11.79\%$
                        & \texttt{sendfile()}     & $10.75\%$ \\

  {\bf NGINX - client}  & \texttt{epoll\_ctl()}   & $29.65\%$
                        & \texttt{connect()}      & $18.95\%$
                        & \texttt{write()}        & $13.47\%$
                        & \texttt{read()}         & $10.70\%$ \\

  {\bf PostMark}        & \texttt{write()}        & $52.27\%$
                        & \texttt{read()}         & $36.67\%$
                        & \texttt{unlink()}       & \phantom{0}$6.05\%$
                        & \texttt{openat()}       & \phantom{0}$2.34\%$ \\
  \hline
\end{tabular}
\end{footnotesize}
\end{table*}

To assess the overhead of {\System} on real-world programs, we used the
Phoronix Test Suite~\cite{benchmark:PTS} v9.8.0 (Nesodden).
Phoronix is an open-source automated benchmarking suite with over 450
different test profiles grouped into categories such as
disk, network, processor, graphics, and system.

We chose the set of tests used to track the
performance of the mainline Linux
kernel~\cite{benchmark:PTS:LinuxRegression}.  We omitted stress
tests, which resemble the LMBench microbenchmarks
of Section~\ref{section:eval:microbenchmarks:iskios}.
Our chosen tests cover
(a) web servers (\texttt{Apache} and \texttt{NGNIX});
(b) the build toolchains for Apache
(\texttt{Build Apache}) and PHP 7 (\texttt{Build PHP});
(c) encryption software (\texttt{Cryptsetup-sha215},
\texttt{Cryptsetup-whirlpool}, and \texttt{OpenSSL});
(d) the PHP interpreter (\texttt{PHPBench}) and LuaJIT
compiler (\texttt{LuaJIT});
(e) the \textit{PostgreSQL} object-relational database search
(\texttt{PGBench}) and a protein sequence database searched via Hidden
Markov Models (\texttt{HMMer});
(f) web and mail server workloads (\texttt{Ebizzy} and \texttt{PostMark});
(g) filesystem and disk I/O workloads (\texttt{FS-Mark} and \texttt{CompileBench});
(h) video encoding (\texttt{SVT-AV1} and \texttt{SVT-HEVC});
(i) audio encoding (\texttt{FLAC} and \texttt{MP3});
(j) file compression (\texttt{Parallel BZIP2});
(k) game engines for chess (\texttt{Crafty}, \texttt{Stockfish}) and Connect-4
(\texttt{Fhourstones});
(l) computation-heavy prime number generation
(\texttt{Primesieve}), Poisson equation solving (\texttt{Himeno}), and
matrix multiplication (\texttt{PolyBench-C});
(m) system boot-up performance
(\texttt{Systemd Kernel});
(n) 3D computer graphics raytracing (\texttt{C-Ray}) and
rendering (\texttt{TTSIOD 3D Renderer}); and
(o) version control (\texttt{Git}).

Phoronix keeps running a benchmark until the relative standard deviation (RSD)
falls below a specific threshold
($3.5\%$ by default) or a maximum number of runs (15 in our experiments) is
exhausted.  Table~\ref{table:eval:phoronix} presents {\System}'s
overhead on each benchmark. The second column shows the
metric used by each benchmark and the result on the \texttt{vanilla}
kernel (i.e., our baseline).
Columns 3--6 show the percentage overhead of each configuration over the
baseline kernel.
The final row shows the geometric mean, across all applications, of the
change in reported metric ($(\prod_i (overhead_{i} + 100))^{1/n}-100$).  As
different applications use different metrics, this average should be
taken as merely suggestive; nonetheless, overhead of less than 5\% when
running on the full version of \System, with XOM and a race-free
protected shadow stack, strikes us as a very strong result.

For \texttt{PKK} alone, we observe a maximum overhead of
$\sim$17\% for \texttt{CompileBench} and \texttt{Apache}, followed by
$\sim$14\% for \texttt{NGINX} and $\sim$9\% for \texttt{PostMark}.
Table~\ref{table:eval:strace} presents the top four system calls with respect
to system time---i.e., CPU time in the kernel---for each of
these four macrobenchmarks, as reported by
\texttt{strace}. \texttt{CompileBench} tests file system performance by
simulating kernel builds (creating,
compiling, patching, stating, and reading kernel trees).
We examined the initial creation of 1000 1~MB files.
Table~\ref{table:eval:strace} shows that
$\sim$61\% of system time is spent on \texttt{write()},
followed by \texttt{openat()}, \texttt{fstat()}, and \texttt{close()}. As
Section~\ref{section:eval:microbenchmarks} explains, {\System}'s
\texttt{PKK} affects these calls the most, as their
total execution time is small.
Since \texttt{CompileBench} uses extremely small services
frequently, the overhead of \texttt{PKK} is higher.
\texttt{PostMark} simulates the behavior of mail servers and thus
of small-file testing (file sizes of 5--512\,KB).
Table~\ref{table:eval:strace} shows that \texttt{PostMark} spends most of its
system time ($\sim$52\%) in \texttt{write()}, followed by
\texttt{read()}, \texttt{unlink()}, and \texttt{openat()}.
As in \texttt{CompileBench}, frequent use of such short kernel calls
leads to higher overheads.

For \texttt{NGINX} and \texttt{Apache}, we used
\texttt{strace} to examine both client and server system calls.
Table~\ref{table:eval:strace} shows that the \texttt{Apache} server
spends $91\%$ of its system time in \texttt{futex()},
followed by \texttt{read()}, \texttt{epoll\_}\linebreak[1]\text{wait()},
and \texttt{accept4()}.
The \texttt{NGINX} server spends roughly the same amount of system time among
\texttt{close()}, \texttt{recvfrom()}, \texttt{writev()}, and
\texttt{sendfile()}.
Table~\ref{table:eval:strace} also shows that clients in both cases
spend most of their system time in \texttt{epoll\_ctl()}, \texttt{connect()},
\texttt{write()}, and \texttt{read()}. This is expected, as the two
benchmarks share the
\texttt{Apache Bench} (\texttt{ab}) client.
Calls to \texttt{futex()} are used
for synchronization and are typically very brief.
The \texttt{connect()} system call is used to
initiate a connection over a socket and \texttt{accept4()} to accept
such connections. The \texttt{recvfrom()} and \texttt{sendfile()} system calls
are used to receive and send data over a socket, respectively.
As Section~\ref{section:eval:microbenchmarks} explains,
\texttt{PKK} incurs significant
overhead on such services, especially when running TCP
(as both \texttt{NGINX} and \texttt{Apache} do).
The \texttt{epoll\_ctl()} and \texttt{epoll\_wait()} system calls are used to
synchronize communication over a set of file descriptors, and much like
\texttt{select()} (discussed in
Sec.~\ref{section:eval:microbenchmarks}),
they suffer high overhead when the number of descriptors is high.

Within the bounds of experimental error, {\System}'s \texttt{XOM}
implementation incurs no overhead beyond that of \texttt{PKK}.
The fully-optimized shadow stack implementation
(\texttt{pkk+}\linebreak[1]\texttt{ss-}\linebreak[1]\texttt{swo-}%
\linebreak[1]\texttt{inline}), as expected, increases the overheads for
macrobenchmarks, such as web servers and file-system workloads, that spend
significant time in the kernel.  The maximum overhead across all
benchmarks and configurations is $38.51\%$;
it occurs for the \texttt{Apache} web server when both security defenses are
enabled (\texttt{pkk+xom+ss-swo-inline}).
\texttt{NGINX} is similar (maximum overhead of $37.14\%$).

\subsection{Code Size Overhead}
\label{section:eval:codesize}

\begin{table}[bt]
\caption{{\System} Code Size Overheads}
\label{table:eval:code-size}
\centering
  \begin{footnotesize}
  \begin{tabular}{|c|r|r|}
    \hline
    \textbf{Kernel}   & \textbf{Code Size} & \textbf{Overhead } \\
    \hline
    {\tt \bf vanilla}           & $28.01   \hphantom{0}\mathit{MB}$
                                & $-$ \\
    {\tt \bf pkk}               & $30.2   \hphantom{0}\mathit{MB}$
                                & $7.82\%$ \\
    {\tt \bf pkk+xom}           & $30.2   \hphantom{0}\mathit{MB}$
                                & $7.82\%$ \\
    {\tt \bf pkk+ss}            & $71.5   \hphantom{0}\mathit{MB}$
                                & $155.27\%$ \\
    {\tt \bf pkk+ss-swo}        & $73.75   \hphantom{0}\mathit{MB}$
                                & $163.30\%$ \\
    {\tt \bf pkk+ss-swo-inline} & $135.7  \hphantom{0}\mathit{MB}$
                                & $384.47\%$ \\
    {\tt \bf pkk+xom+ss}        & $71.5   \hphantom{0}\mathit{MB}$
                                & $155.27\%$ \\
    {\tt \bf pkk+xom+ss-swo}    & $73.75  \hphantom{0}\mathit{MB}$
                                & $163.30\%$ \\
    {\tt \bf pkk+xom+ss-swo-inline} & $135.7  \hphantom{0}\mathit{MB}$
                                    & $384.47\%$ \\
    \hline
  \end{tabular}
  \end{footnotesize}
\end{table}

To see {\System}'s impact on code size, we
measured the {\tt .text} section in the final binary for each
configuration (including all the loaded modules).
Results appear in Table~\ref{table:eval:code-size}.
{\System}'s \texttt{PKK} instrumentation has relatively little
impact---$7.82\%$---as it only adds a few lines of code in the kernel
entry/exit path and two instructions for every indirect branch.

As expected, {\System}'s \texttt{XOM} does not increase the code size
relative to \texttt{PKK}. In contrast, SFI-based XOM approaches,
which instrument every load instruction, have higher memory overheads:
existing implementations of execute-only memory for user-level
applications e.g., LR\textsuperscript{2}~\cite{LR2:NDSS16} and
uXOM~\cite{uXOM:SEC19}, report code-size increases of $10\%$ to $50\%$.
Unfortunately, kR\textasciicircum X~\cite{kRX:TOPS19}, the only SFI-based approach
for \texttt{XOM} in the Linux kernel, does not report code-size overheads.

{\System}'s unoptimized shadow stack implementation incurs a $155\%$
increase in code segment size. This is expected, as our shadow stack
compiler pass adds 25 instructions to each callsite
(Listing~\ref{lst:ss-callsite}) and one instruction
to each function epilogue (Listing~\ref{lst:ss-epilogue}).
The shadow-write optimization (\texttt{SWO}) adds two more instructions per
callsite (Listing~\ref{lst:ss-callsite}), for another $\sim$8\% overhead
compared to the unoptimized shadow stack and $163\%$ compared to
our baseline.
\texttt{XOM} again adds nothing more to the
code size of the shadow stack implementation.
Finally, increasing Clang's inlining threshold to 8000
adds roughly $385\%$ overhead compared to the baseline, which is also the
cumulative overhead when both security defenses are enabled.
While this increase is high, the absolute size of the kernel along with the
loaded modules in {\System} is less than $136$~MB, which we consider
entirely reasonable given the added security and the fact that modern
desktop and server machines come with 8~GB or more of physical memory.

\section{Related Work}
\label{section:related}

We discuss three areas of related work:
intra-address-space memory isolation,
execute-only memory defenses, and methods to protect the
integrity of return addresses.

\subsection{Intra-address-space Isolation}

\textbf{Address-based Isolation}
Several research efforts~\cite{KCoFI:Oakland14, kRX:TOPS19, PittSFIeld:2005,
Datashield:ASIACCS17, NaCL:CACMJan10}
employ \emph{software fault isolation} (SFI)~\cite{SFI:SOSP93,SFI2:UsenixSec10} to
protect sensitive data from untrusted
code. Typically, these approaches add a run-time check
on each memory access to ensure that the target address is not within
the protected region. Additionally, they enforce some form of
\emph{control-flow integrity} (CFI)~\cite{CFI:TISSEC09} to ensure that the SFI
instrumentation is not bypassed.
As the protection overhead is proportional to the
complexity of the code performing the checks, solutions using SFI typically
1) protect only a single memory region, and 2) place
pages belonging to the same protection domain contiguously within the virtual
address space~\cite{KCoFI:Oakland14, kRX:TOPS19, Datashield:ASIACCS17,
NaCL:CACMJan10}.
Unfortunately, this approach restricts the number of protected domains and
requires significant engineering effort
(memory allocator modifications in particular~\cite{kRX:TOPS19}).
In contrast, our PKK supports up to 8 distinct protection domains and
permits pages in different domains to be located anywhere within the virtual
address space without incurring additional performance loss.

\textbf{Domain-based Isolation}
Other approaches~\cite{Hodor:ATC19, ERIM:UsenixSec19, MonGuard:EuroSec20,
Intra-Unikernel:VEE20, UnderBridge:ATC20, SEIMI:Oakland20, LVDs:VEE20} use
hardware isolation primitives to
create \emph{protection domains} in which various executable components are
granted asymmetric access to sensitive data in memory.
Hodor~\cite{Hodor:ATC19},
ERIM~\cite{ERIM:UsenixSec19}, and MonGuard~\cite{MonGuard:EuroSec20}
use PKU to provide efficient intra-process isolation for user-space programs.
Like {\System}, these systems use memory protection keys.
However, {\System} provides intra-kernel
isolation and, consequently, solves several challenges such as
retrofitting PKU for protecting kernel memory while maintaining
SMAP and SMEP\@.
Sung et al.~\cite{Intra-Unikernel:VEE20} leverage Intel PKU to provide
intra-address-space isolation for the RustyHermit~\cite{RustyHermit:PLOS19}
unikernel. A RustyHermit~\cite{RustyHermit:PLOS19} unikernel instance consists
of a single application, statically linked against a minimal library OS written
in Rust~\cite{RustyHermit:PLOS19}; application and kernel share the same
\emph{user address space} (i.e., pages with U/S bit set), and they both
execute in \emph{kernel mode}. Sung et al.'s~\cite{Intra-Unikernel:VEE20}
proposed mechanism places the application in a different protection
domain from the kernel, and it isolates unsafe Rust code blocks
that handle low-level hardware operations from the rest of the kernel.
As all code uses user-space memory pages, the use of Intel PKU
is straightforward. Since the goal is to increase fault tolerance within
a single domain of trust, Sung et al.~\cite{Intra-Unikernel:VEE20} do not
prevent reconfiguration of the {\tt pkru} register.
Even if they did, their system
would not need to differentiate between \texttt{wrpkru} instructions
in user and kernel space.
In contrast, {\System} retrofits Linux's
existing user/kernel isolation mechanism to use
protection keys. In doing so, {\System} takes additional
measures to ensure that user-space application code cannot misuse the
\texttt{wrpkru} instruction to break {\System}'s isolation mechanism.
UnderBridge~\cite{UnderBridge:ATC20}, developed concurrently to our work,
restructures microkernel OSes by moving user-space system servers to
kernel space; it uses
PKU in conjunction with \emph{Kernel Page Table Isolation
(KPTI)}~\cite{KPTI:LKML} to isolate system servers running in kernel mode.
While compatible with KPTI, our PKK mechanism does
not require it, and
therefore PKK does not incur KPTI's additional overhead~\cite{KPTI-cost}.

Other x86 hardware primitives can be used for
intra-address-space isolation.
Several works~\cite{LVDs:VEE20,SkyBridge:EuroSys19,Hodor:ATC19,SeCage:CCS15,
MemSentry:EuroSys17,xMP:SP20} use
Intel's \texttt{\slshape VMFUNC}~\cite{IntelArchManual21} to switch between preset
\emph{Extended Page Tables (EPTs)} that correspond to different
protection domains.
Among them, \mbox{SkyBridge}~\cite{SkyBridge:EuroSys19} uses \texttt{VMFUNC} to
improve the latency of IPCs in a microkernel setting,
LVDs~\cite{LVDs:VEE20} use \texttt{VMFUNC} to isolate device drivers from
the core Linux kernel, and xMP~\cite{xMP:SP20} uses the same mechanism to
protect sensitive user and kernel data from data-only attacks.
\texttt{VMFUNC} requires that these systems execute
under a hypervisor, which incurs non-trivial virtualization overheads on
normal execution.
Conversely, {\System} uses no hypervisor,
keeping its overheads during normal execution low.
Additionally, as \texttt{VMFUNC} is 3--5$\times$ slower than
\texttt{wrpkru}~\cite{Hodor:ATC19, ERIM:UsenixSec19, SEIMI:Oakland20}, its use
for defenses requiring frequent domain switching (e.g., shadow
stacks) is prohibitive~\cite{ERIM:UsenixSec19, SEIMI:Oakland20}.

SEIMI~\cite{SEIMI:Oakland20} uses
\emph{SMAP}~\cite{IntelArchManual21} to protect sensitive data in user-space
programs. It enables SMAP and maps all pages with the supervisor bit except
for those that store sensitive user data. SEIMI then executes
\emph{all} (trusted and untrusted) user code in privileged mode and allows
trusted components to temporarily disable SMAP to access the protected memory.
To prevent user code from executing privileged instructions,
SEIMI places the system under
the control of a
hypervisor and forces all privileged instructions to cause a VM exit.
Both {\System} and SEIMI make unconventional use
of x86 hardware to efficiently isolate memory.
Unlike SEIMI, {\System} does not require a
hypervisor to protect its isolation mechanism and security defenses.
In addition, SEIMI's use of SMAP no longer protects the kernel from
inadvertently accessing user memory.
In contrast, {\System}'s PKK mechanism still allows user programs to use
PKU (though it does halve the number of user-level keys available).

\subsection{eXecute-Only Memory}

Defenses in user space prevent code pages from being
read~\cite{Readactor:SP15, LR2:NDSS16, HideM:CODASPY15, XnR:CCS14, uXOM:SEC19}
and use code-pointer hiding techniques~\cite{Readactor:SP15, LR2:NDSS16} to
protect against indirect memory disclosure
attacks.
Inspired by their user-space counterparts, recent approaches in the OS
kernel~\cite{KernelDiversification:CNS16, kRX:TOPS19} combine
execute-only memory with kernel diversification to prevent gadgets from
being leaked.
KHide~\cite{KernelDiversification:CNS16} uses
a hypervisor to prevent read accesses to kernel code.
{\System} does not require more privileged software
for its execution, keeping its trusted computing base small
and avoiding unnecessary virtualization overheads.
kR\textasciicircum X~\cite{kRX:TOPS19}
instruments all read instructions with run-time checks to ensure that
they never read the code segment. In its fully-optimized software-based
implementation, kR\textasciicircum X performs roughly the same as {\System}; however,
kR\textasciicircum X must place all
code in a contiguous region, weakening
diversification schemes
such as KASLR~\cite{KASLR:LKML}. To make up for the entropy loss, kR\textasciicircum X
re-arranges code in the protected region using function permutation and
basic block reordering~\cite{kRX:TOPS19}.
In contrast, {\System} does not break the memory layout, provides more
flexibility (i.e., supports more than one protected area
in memory), and preserves the protection guarantees of existing
randomization schemes. A faster implementation of kR\textasciicircum X exists but relies on
Intel MPX which is now deprecated~\cite{MPXDeprecated}.

\subsection{Return Address Protection}

Stack canaries~\cite{StackGuard:UsenixSec98,Propolice:GCCplugin,
Stackguard:GCCplugin} and systems that encrypt return
addresses~\cite{PointGuard:UsenixSec03, PAX:RAP, kRX:TOPS19, G-Free:ACSAC10,
PACitup:UsenixSec19, ARM-PA:Qualcomm}
can detect return address corruption on the stack but
are susceptible to memory disclosure~\cite{PointGuard:UsenixSec03}, brute-force~\cite{PACitup:UsenixSec19},
signing-key forgery~\cite{KeyForgery:Google}, and substitution
attacks~\cite{PACitup:UsenixSec19, ARM-PA:Qualcomm}.
LLVM Safestack~\cite{LLVMSafestack} stores return addresses on a
second separate stack which it stores in a randomized location.
Other defenses create a separate (shadow) stack in memory and, on each
function call, store a copy of the return address to the shadow
stack~\cite{StackShield:GCCplugin, RAD:ICDCS01, ShadowStack:ASIACCS15,
SoKShiningLight:SP19}. Shadow stacks may or may not be write-protected.
Systems deploying the latter (e.g., Shadesmar~\cite{SoKShiningLight:SP19})
place the shadow stack at a random location in memory like the LLVM
Safestack~\cite{LLVMSafestack}.
While these solutions make attacks more difficult---the base
address needs to be guessed correctly and, in the case of shadow stacks, two
return addresses must be corrupted in memory---they are still susceptible to
information leaks and memory corruption attacks~\cite{DualStack:ASIACCS18,
SoKShiningLight:SP19, LLVMSafestack, MissingPointer:SP15}.

Write-protected shadow stacks (e.g., Read-Only RAD~\cite{RAD:ICDCS01}), if
implemented correctly to be thread-safe, offer the strongest protection for
return addresses.
Prior works~\cite{SoKShiningLight:SP19, SEIMI:Oakland20} examine the
performance overheads of various hardware mechanisms to preserve the
integrity of a user-space shadow stack. Burow et
al.~\cite{SoKShiningLight:SP19} report a $12.12\%$ geo\-mean overhead over
the SPEC CPU2006~\cite{SPEC:CPU2006} benchmark suite when using
Intel MPX~\cite{IntelArchManual21} and a $61.18\%$ geomean overhead when
using Intel PKU. Wang et al.~\cite{SEIMI:Oakland20} report a $14.57\%$ geomean
overhead over the same set of benchmarks for the MPX-based implementation of a
shadow stack, and a $21.08\%$ geomean overhead for the PKU-based shadow stack.
Unfortunately, Intel MPX is deprecated and will be unavailable in
future processors~\cite{MPXDeprecated}.
SEIMI's \texttt{SMAP}-based solution incurs $12.49\%$ geomean
overhead but, as noted above, requires hardware virtualization, adding
significant overheads to the entire system---$68.37\%$ and $33.56\%$ geomean
overhead on process- and filesystem-related kernel operations, respectively;
it also no longer supports \texttt{SMAP}'s original purpose.
In contrast, {\System} uses hardware that is available in all
processors following the Intel Skylake generation, does not require any
virtualization support, keeping the overheads of kernel operations low,
maintains half of the protection keys for userspace protection, and incurs
only $22\%$ geomean overhead on the LMBench latency microbenchmarks while
providing robust race-free protection to kernel shadow stacks.

Microsoft's \emph{Hardware-enforced Stack
  Protection}~\cite{Microsoft-CET} is the first security mechanism to
use the long-awaited Intel \emph{Control-Flow Enforcement Technology
  (CET)} for return addresses.
On each function call, CET stores the return address on both the
program stack and a hardware-enforced read-only shadow stack.
On return, it compares the two addresses and raises an exception
in case of a mismatch.  Unfortunately, at the time of this writing, CET is
available only on certain \emph{Tiger Lake} mobile and embedded processors.

\section{Discussion and Future Work}
\label{section:future}

\textbf{Side Channels}
As Section~\ref{section:pkk:side} discusses, PKK introduces a side channel
through which user code may infer locations in the kernel code segment. An
alternative design for PKK would be to only set the U/S bit on kernel
data pages and let kernel code reside in supervisor-mode pages. Such a
design would offer equivalent security guarantees
and would avoid the additional overheads of our SMEP
instrumentation, but it would not support kernel XOM\@.
Note that enabling shadow stack support in our current prototype narrows
the window of side channel vulnerability by introducing vetted
\texttt{wrpkru} instructions at each function call site.

\textbf{Intel PKS}
When available, we could leverage Intel's supervisor protection
keys~\cite{IntelArchManual21} and eliminate the need for SMEP emulation.
Such a system would still require additional code after {\tt wrmsr}
instructions to prevent code reuse attacks from disabling protection
(as {\System} does for {\tt wrpkru} instructions).
Our shadow stack optimizations should be
compatible with PKS\@.

\textbf{Register Allocation}
Given better integration with local register allocation, \System\ could
often avoid spilling the three registers used to instrument  call sites.
Inter-procedural register allocation could similarly avoid spilling
the return address to the shadow stack in many cases, further reducing
the frequency of {\tt wrpkru} instructions.

\textbf{Fine-grained Intra-kernel Isolation}
Additional future work might use PKK to protect sensitive kernel data
regions---e.g., process control blocks (PCBs), credential structures,
and interrupt vector tables---against unauthorized accesses.  We also
hope to investigate the use of PKK to better isolate kernel components.
Previous work in this area~\cite{NestedKernel:ASPLOS15} relies on
expensive serializing instructions and can only protect data
integrity; PKK offers a potentially faster
mechanism that can ensure both integrity and confidentiality.

\section{Conclusions}
\label{section:concl}

{\System} is, to the best of our knowledge, the first system
to implement race-free write-protected shadow stacks and flexible
(non-entropy-compromising) execute-only
memory for the OS kernel. Shadow stacks protect return addresses from
corruption, and
execute-only memory enables state-of-the-art
leakage-resilient diversification schemes by hiding code from buffer
overread attacks. Unlike previous work, IskiOS imposes no restrictions on
virtual address space layout, allowing the OS to
place kernel stacks and code pages at
arbitrary locations. {\System} achieves these
benefits through a novel use of Intel's PKU for protection inside the OS
kernel.  Our ``PKK''-based implementation of XOM (with
software-emulated SMEP) incurs roughly $12\%$ geomean overhead on the
LMBench
microbenchmarks, relative to the vanilla Linux kernel
and virtually no performance overhead on most real-world applications
(less than 16\%, worst case, on the Phoronix programs used
for Linux regression analysis).  {\System}'s
protected shadow stacks incur about 22\% geomean overhead
on LMBench and less than 5\% (geomean) across the chosen Phoronix
applications (less than 39\%, worst case).

\begin{acks}
The authors thank the anonymous referees and our shepherd, 
Erik van der Kouwe, for their helpful feedback and suggestions.
This work was supported in part by NSF grants
\pagebreak      
CNS-1618213,
CNS-1900803,
and
CNS-1955498, by a Google Faculty Research award,
and by
ONR Award N00014-17-1-2996.
\end{acks}

\bibliographystyle{ACM-Reference-Format}
\bibliography{references}

\appendix


\begin{table*}[tb]
\caption{{\System} run-time overhead ($\%$ over {\tt vanilla} Linux) on
latency microbenchmarks of the LMBench suite.} \label{table:eval:lmbench-lat}
\begin{minipage}{\textwidth}
\centering
\resizebox{\textwidth}{!}{%
\begin{tabular}{ |l|rr|r|r|r|r|r|r|r|r|}
\hline
                              \multirow{2}{*}{\bf Benchmark}
& \multicolumn{2}{c|}{\multirow{2}{*}{\bf vanilla}}
& \multicolumn{2}{c|}{\tt \bf pkk}
& \multicolumn{2}{c|}{\tt \bf pkk+xom}
& \multicolumn{2}{c|}{\tt \bf pkk+ss-swo-inline}
& \multicolumn{2}{c|}{\tt \bf pkk+xom+ss-swo-inline}\\
\cline{4-11}
  & & &Mean $\pm$ SD &Overhead &Mean $\pm$ SD &Overhead &Mean $\pm$ SD &Overhead &Mean $\pm$ SD &Overhead \\
\hline
{\tt \bf syscall()}         & $0.24 \pm 0.00$ &$\mu s$
                            & $0.29 \pm 0.00$ & $20.88\%$
                            & $0.29 \pm 0.00$ & $24.05\%$
                            & $0.29 \pm 0.01$ & $22.98\%$
                            & $0.29 \pm 0.00$ & $21.92\%$\\
{\bf open()/close()}        & $1.30 \pm 0.00$ & $\mu s$
                            & $1.48 \pm 0.00$ & $14.12\%$
                            & $1.48 \pm 0.01$ & $14.40\%$
                            & $1.85 \pm 0.01$ & $42.61\%$
                            & $1.84 \pm 0.01$ & $41.67\%$\\
{\bf read()}                & $0.29 \pm 0.00$ & $\mu s$
                            & $0.35 \pm 0.00$ & $20.80\%$
                            & $0.35 \pm 0.00$ & $19.80\%$
                            & $0.39 \pm 0.00$ & $31.99\%$
                            & $0.39 \pm 0.00$ & $31.87\%$\\
{\bf write()}               & $0.27 \pm 0.00$ & $\mu s$
                            & $0.33 \pm 0.00$ & $24.66\%$
                            & $0.32 \pm 0.00$ & $19.84\%$
                            & $0.32 \pm 0.01$ & $18.21\%$
                            & $0.31 \pm 0.00$ & $17.98\%$\\
 {\bf select(10 fds)}       & $0.39 \pm 0.00$ & $\mu s$
                            & $0.46 \pm 0.00$ & $17.26\%$
                            & $0.47 \pm 0.00$ & $20.15\%$
                            & $0.47 \pm 0.00$ & $18.60\%$
                            & $0.45 \pm 0.00$ & $15.42\%$\\
{\bf select(100 TCP fds)}   & $3.34 \pm 0.04$ & $\mu s$
                            & $3.65 \pm 0.04$ & $9.12\%$
                            & $3.72 \pm 0.15$ & $11.21\%$
                            & $3.77 \pm 0.21$ & $12.64\%$
                            & $3.60 \pm 0.15$ & $7.65\%$\\
{\bf select(200 fds)}       & $3.07 \pm 0.11$ & $\mu s$
                            & $3.16 \pm 0.00$ & $2.97\%$
                            & $3.17 \pm 0.00$ & $3.36\%$
                            & $3.26 \pm 0.03$ & $6.24\%$
                            & $3.23 \pm 0.01$ & $5.45\%$\\
{\bf stat()}                & $0.65 \pm 0.01$ & $\mu s$
                            & $0.78 \pm 0.00$ & $20.11\%$
                            & $0.76 \pm 0.00$ & $18.01\%$
                            & $0.83 \pm 0.01$ & $28.52\%$
                            & $0.90 \pm 0.00$ & $39.75\%$\\
{\bf fstat()}               & $0.29 \pm 0.00$ & $\mu s$
                            & $0.36 \pm 0.00$ & $23.83\%$
                            & $0.36 \pm 0.00$ & $23.41\%$
                            & $0.39 \pm 0.00$ & $35.96\%$
                            & $0.45 \pm 0.00$ & $54.81\%$\\
{\bf fcntl()}               & $1.93 \pm 0.03$ & $\mu s$
                            & $2.00 \pm 0.03$ & $4.01\%$
                            & $2.05 \pm 0.03$ & $6.37\%$
                            & $2.24 \pm 0.02$ & $16.57\%$
                            & $2.24 \pm 0.02$ & $16.27\%$\\
{\bf mmap()/munmap()}       & $71.90 \pm 0.88$ & $\mu s$
                            & $76.60 \pm 1.17$ & $6.54\%$
                            & $76.20 \pm 0.63$ & $5.98\%$
                            & $88.00 \pm 1.76$ & $22.39\%$
                            & $86.90 \pm 1.52$ & $20.86\%$\\
{\bf fork()+exit()}         & $328.10 \pm 12.70$ & $\mu s$
                            & $352.94 \pm 5.62$ & $7.57\%$
                            & $355.22 \pm 5.67$ & $8.27\%$
                            & $418.47 \pm 7.28$ & $27.54\%$
                            & $406.75 \pm 8.94$ & $23.97\%$ \\
{\bf fork()+execve()}       & $335.15 \pm 4.63$ & $\mu s$
                            & $369.95 \pm 5.40$ & $10.38\%$
                            & $363.33 \pm 5.73$ & $8.41\%$
                            & $430.35 \pm 7.39$ & $28.41\%$
                            & $420.13 \pm 7.18$ & $25.36\%$\\
{\bf fork()+/bin/sh}        & $2744.75 \pm 8.81$ & $\mu s$
                            & $2953.10 \pm 3.04$ & $7.59\%$
                            & $2967.65 \pm 8.68$ & $8.12\%$
                            & $3356.80 \pm 8.86$ & $22.30\%$
                            & $3308.80 \pm 13.49$ & $20.55\%$\\
{\bf sigaction()}           & $0.28 \pm 0.00$ & $\mu s$
                            & $0.34 \pm 0.00$ & $21.35\%$
                            & $0.36 \pm 0.00$ & $27.11\%$
                            & $0.36 \pm 0.00$ & $26.90\%$
                            & $0.36 \pm 0.00$ & $26.94\%$ \\
{\bf Signal Delivery}       & $1.05 \pm 0.00$ & $\mu s$
                            & $1.13 \pm 0.01$ & $7.30\%$
                            & $1.15 \pm 0.00$ & $8.81\%$
                            & $1.41 \pm 0.01$ & $34.13\%$
                            & $1.41 \pm 0.00$ & $33.68\%$ \\
{\bf Protection Fault}      & $0.71 \pm 0.00$ & $\mu s$
                            & $0.76 \pm 0.01$ & $6.48\%$
                            & $0.77 \pm 0.01$ & $8.59\%$
                            & $0.60 \pm 0.00$ & $-15.75\%$
                            & $0.61 \pm 0.01$ & $-14.10\%$\\
{\bf Page Fault}            & $0.22 \pm 0.00$ & $\mu s$
                            & $0.24 \pm 0.00$ & $6.01\%$
                            & $0.23 \pm 0.00$ & $5.28\%$
                            & $0.24 \pm 0.01$ & $8.14\%$
                            & $0.24 \pm 0.00$ & $6.82\%$\\
{\bf Pipe I/O}              & $6.76 \pm 0.10$ & $\mu s$
                            & $6.96 \pm 0.06$ & $2.92\%$
                            & $6.96 \pm 0.09$ & $2.99\%$
                            & $7.48 \pm 0.11$ & $10.64\%$
                            & $7.44 \pm 0.06$ & $10.11\%$\\
{\bf Unix Socket I/O}       & $6.56 \pm 0.01$ & $\mu s$
                            & $7.42 \pm 0.04$ & $13.14\%$
                            & $7.31 \pm 0.04$ & $11.58\%$
                            & $8.24 \pm 0.05$ & $25.75\%$
                            & $8.15 \pm 0.05$ & $24.35\%$\\
{\bf TCP Socket I/O}        & $13.08 \pm 0.18$ & $\mu s$
                            & $14.02 \pm 0.12$ & $7.16\%$
                            & $14.07 \pm 0.05$ & $7.55\%$
                            & $16.87 \pm 0.07$ & $28.94\%$
                            & $16.93 \pm 0.14$ & $29.42\%$\\
{\bf UDP Socket I/O}        & $10.39 \pm 0.08$ & $\mu s$
                            & $11.25 \pm 0.11$ & $8.29\%$
                            & $11.52 \pm 0.12$ & $10.91\%$
                            & $14.16 \pm 0.07$ & $36.30\%$
                            & $14.48 \pm 0.43$ & $39.40\%$\\
{\bf Context switch}        & $2.58 \pm 0.06$ & $\mu s$
                            & $2.58 \pm 0.04$ & $0.04\%$
                            & $2.57 \pm 0.05$ & $-0.35\%$
                            & $2.82 \pm 0.06$ & $9.18\%$
                            & $2.86 \pm 0.03$ & $10.77\%$\\
\hline
\multicolumn{3}{|c|}{\bf Geometric Mean Overhead} & &
    $11.18\%$ & & $11.67\%$ & & $21.01\%$ & & $22.33\%$ \\
\hline
\end{tabular}
}

\end{minipage}
\end{table*}

\begin{table*}[tb]
\caption{{\System} run-time overhead ($\%$ over {\tt vanilla} Linux) on
bandwidth microbenchmarks of the LMBench suite.} \label{table:eval:lmbench-bw}
\begin{minipage}{\textwidth}
\centering
\resizebox{\textwidth}{!}{%
\begin{tabular}{ |l|rr|r|r|r|r|r|r|r|r|}
 \hline
                              \multirow{2}{*}{\bf Benchmark}
& \multicolumn{2}{c|}{\multirow{2}{*}{\bf vanilla}}
& \multicolumn{2}{c|}{\tt \bf pkk}
& \multicolumn{2}{c|}{\tt \bf pkk+xom}
& \multicolumn{2}{c|}{\tt \bf pkk+ss-swo-inline}
& \multicolumn{2}{c|}{\tt \bf pkk+xom+ss-swo-inline}\\
\cline{4-11}
  & & &Mean $\pm$ SD &Overhead &Mean $\pm$ SD &Overhead &Mean $\pm$ SD &Overhead &Mean $\pm$ SD &Overhead \\
\hline
{\bf Pipe I/O}              & $2879.34 \pm 89.88$   & $MB/s$
                            & $2760.65 \pm 64.33$   & $4.12\%$
                            & $2819.08 \pm 73.25$   & $2.09\%$
                            & $2892.75 \pm 55.22$   & $-0.47\%$
                            & $2800.93 \pm 73.45$   & $2.72\%$\\
{\bf Unix Socket I/O}       & $5903.98 \pm 109.51$  & $MB/s$
                            & $6145.74 \pm 346.57$  & $-4.09\%$
                            & $6323.30 \pm 269.05$  & $-7.10\%$
                            & $6518.00 \pm 460.98$  & $-10.40\%$
                            & $6901.41 \pm 61.81$   & $-16.89\%$\\
{\bf TCP Socket I/O}        & $5033.57 \pm 28.51$   & $MB/s$
                            & $4889.97 \pm 44.86$   & $2.85\%$
                            & $4874.17 \pm 58.14$   & $3.17\%$
                            & $4379.98 \pm 29.15$   & $12.98\%$
                            & $4354.83 \pm 56.52$   & $13.48\%$\\
{\bf File I/O}              & $8155.58 \pm 32.22$   & $MB/s$
                            & $8704.69 \pm 411.54$  & $-6.73\%$
                            & $8287.26 \pm 529.07$  & $-1.61\%$
                            & $8247.62 \pm 30.08$   & $-1.13\%$
                            & $9240.31 \pm 473.58$  & $-13.30\%$\\
 {\bf MMapped File I/O}     & $18769.69 \pm 115.36$ & $MB/s$
                            & $19761.46 \pm 270.87$ & $-5.28\%$
                            & $18532.37 \pm 86.70$  & $1.26\%$
                            & $18739.26 \pm 930.41$ & $0.16\%$
                            & $20225.81 \pm 38.04$  & $-7.76\%$\\
\hline
\multicolumn{3}{|c|}{\bf Geometric Mean Overhead} & & $-1.93\%$ & & $-0.51\%$
      & & $-0.05\%$ & & $-4.98\%$ \\
\hline
\end{tabular}
}
\end{minipage}
\end{table*}

\section{Details of LMBench Experiments}
\label{section:appendix-a}

This appendix includes detailed results from our experiments for interested
readers.

Tables~\ref{table:eval:lmbench-lat} and~\ref{table:eval:lmbench-bw}
show results for our LMBench experiments.
In each, the
second column shows
the arithmetic mean and standard deviation (SD) for ten runs of each latency
and bandwidth microbenchmark on the unmodified Linux kernel
(i.e., our baseline).
Similarly, columns 3--6 show the arithmetic mean and standard deviation, as
well as the percentage overhead, of each configuration over the baseline
kernel. The last row of each table
shows the geometric mean
(in reported metric ($(\prod_i (overhead_{i} + 100))^{1/n}-100$))
overhead across latency and bandwidth microbenchmarks,
respectively, for each {\System} configuration.

\begin{figure}[hb]
  \includegraphics[width=\columnwidth]{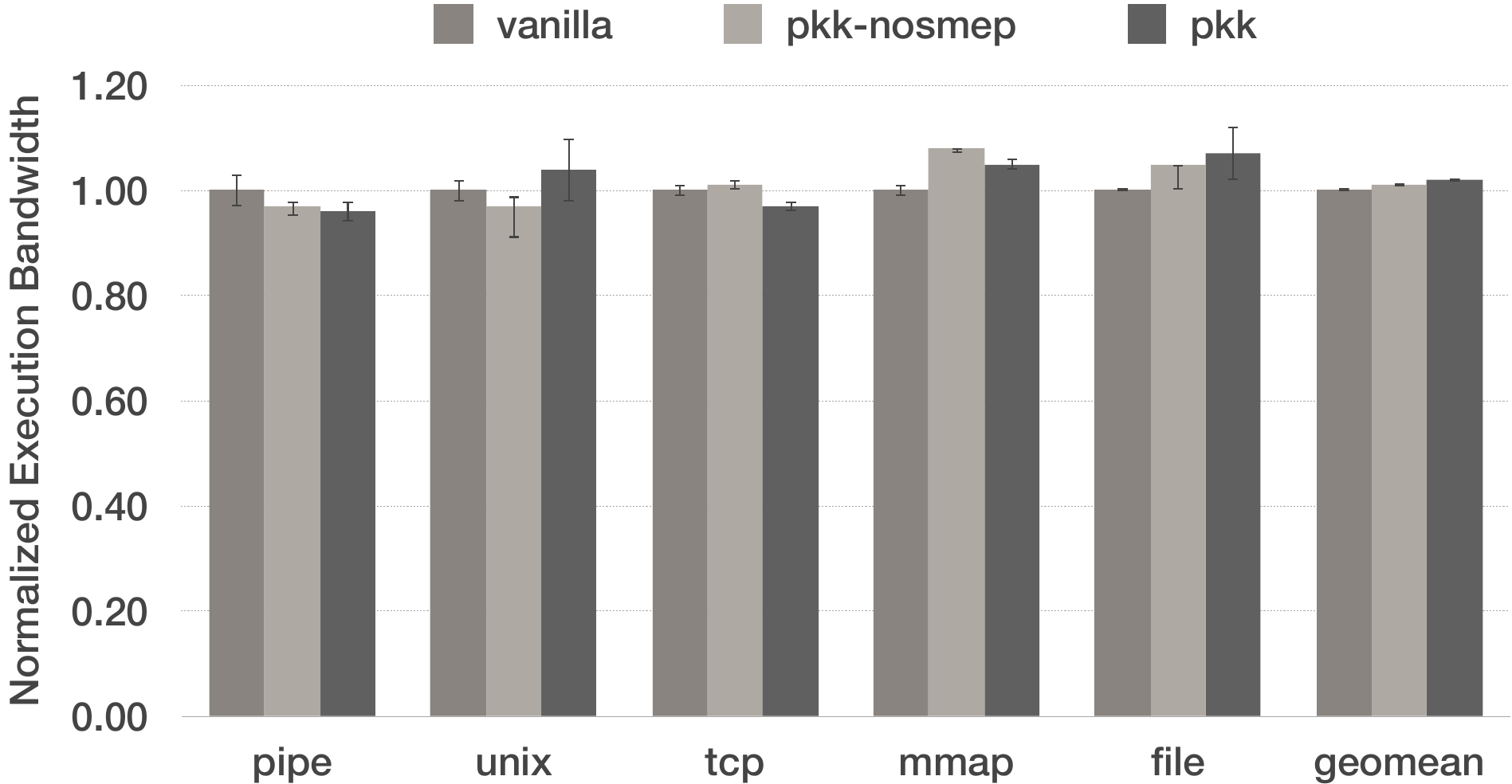}
  \caption{Breakdown of \texttt{PKK} overhead (over {\tt vanilla} Linux)
  on LMBench bandwidth microbenchmarks.
  Columns labeled as \texttt{pkk-nosmep} represent {\System}'s
  \texttt{PKK} implementation with the \texttt{SMEP} LLVM pass disabled.
  }
  \label{fig:pkk-eval:bw}
\end{figure}

\begin{figure*}[b]
  \includegraphics[width=.9\textwidth]{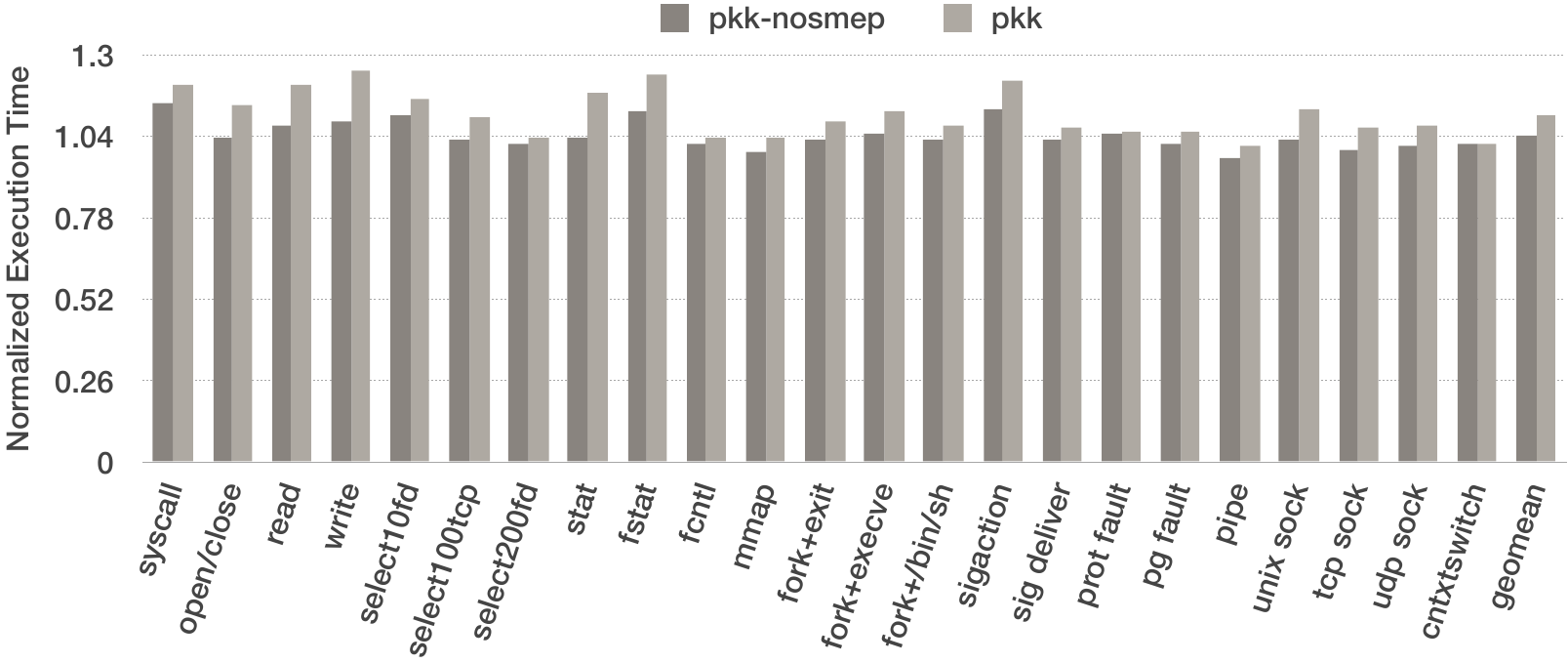}
  \caption{Breakdown of {\System}'s \texttt{PKK} overhead (over {\tt vanilla}
  Linux) on LMBench latency
  microbenchmarks. Columns labeled as \texttt{pkk-nosmep} represent
  {\System}'s \texttt{PKK} implementation with the \texttt{SMEP} LLVM pass
  disabled.}
  \label{fig:pkk-eval:latency}
\end{figure*}

Figures~\ref{fig:pkk-eval:bw} and~\ref{fig:pkk-eval:latency} show the
detailed overhead of using PKK with and without our SMEP
instrumentation on the LMBench bandwidth and latency experiments,
respectively.

\end{document}